\documentclass[aps,prc,twocolumn,superscriptaddress,showpacs,floatfix,nofootinbib]{revtex4-1}
\usepackage{amsmath,graphicx,color}
\usepackage[colorlinks]{hyperref}
\usepackage{tikz}
\usetikzlibrary{patterns}

\begin{document}

\title{Multiplicity fluctuations and correlations in $5.02$~TeV p+Pb collisions at zero impact parameter}

\author{Mathis Pepin}
\affiliation{Institut de physique th\'eorique, Universit\'e Paris Saclay, CNRS, CEA, F-91191 Gif-sur-Yvette, France}
\author{Peter Christiansen}
\affiliation{Lund University Department of Physics, Division of Particle Physics, Lund, Sweden}
\author{St\'ephane Munier}
\affiliation{CPHT, CNRS, \'Ecole polytechnique, IP Paris, F-91128 Palaiseau, France}
\author{Jean-Yves Ollitrault}
\affiliation{Institut de physique th\'eorique, Universit\'e Paris Saclay, CNRS, CEA, F-91191 Gif-sur-Yvette, France} 
\date{\today}

\begin{abstract}
We present a Bayesian method to reconstruct event-by-event multiplicity fluctuations and rapidity correlations in p+Pb collisions at zero impact parameter from minimum-bias data, without assuming any model of the collision dynamics. 
We test it on Monte Carlo simulations with the Angantyr model, then apply it to ATLAS data on the distribution of charged multiplicity and transverse energy in p+Pb collisions at $\sqrt{s_{\rm NN}}=5.02$~TeV. 
Fluctuations in $b=0$ collisions are quantum fluctuations which originate mostly from the proton wave function, and therefore have the potential to constrain the subnucleonic structure of the proton. 
The Angantyr model is found to overestimate fluctuations. 
In addition, we find that as the rapidity increases (towards the Pb-going side), not only the multiplicity density increases, but also its relative event-by-event fluctuation.
This counter-intuitive phenomenon is also observed in simulations with Angantyr, and with the QCD dipole model, where its origin can be traced back to the branching process through which gluons are produced. 
\end{abstract}
\maketitle

\section{Introduction}
It came as a surprise, a decade ago, that central proton-nucleus collisions at the Large Hadron Collider (LHC) create a fluid, despite the tiny size of the collision volume~\cite{Bozek:2011if,ALICE:2012eyl,ATLAS:2013jmi,CMS:2015yux,CMS:2019wiy,Weller:2017tsr}. 
The formation of a fluid implies in particular some equilibration process, which is known to erase the local memory of initial conditions.
One might therefore think that little can be learned from experimental data about the early stages of a proton-nucleus collision. 
Some memory of the initial conditions does, however, remain through global quantities, which are conserved throughout the history of the fluid.
In particular, the rapidity~\cite{Bjorken:1982qr} and the entropy of the fluid are conserved to a good approximation.
Now, the entropy is proportional to the number of elementary constituents of the fluid, whether they are partons~\cite{Ollitrault:2007du} or hadrons~\cite{Hanus:2019fnc}.
Therefore, the final hadron multiplicity reflects the initial gluon multiplicity at the same rapidity.

We study event-by-event fluctuations of the multiplicity and long-range rapidity correlations. 
There are two sources of fluctuations. 
The first is the variation of impact parameter, $b$, across the sample of events. 
Our goal is to isolate the remaining fluctuations, which are quantum fluctuations.   
The observables we consider are typically the multiplicities in two separate rapidity intervals. 
We reconstruct their variances and mutual correlation in collisions at $b=0$ from minimum-bias data by simple Bayesian inference, without assuming any model of the collision dynamics. 

Event-by-event fluctuations in $b=0$ collisions originate from the wave functions of the colliding proton and nucleus, and from the collision process itself. 
For such an asymmetric system as a p+Pb collision, it seems natural that the fluctuations are likely to originate mostly from the smaller of the two projectiles, i.e., the proton.
While there is a vast literature on fluctuations in the proton wave function on the theory side~\cite{Ralston:1988rb,Heiselberg:1991is,Alvioli:2013vk,Schenke:2014zha,Mantysaari:2016ykx,Bierlich:2016smv,Mantysaari:2022ffw}, observables to constrain them are still scarce~\cite{Mantysaari:2016ykx}, and we will show that proton-nucleus collisions bring new, non-trivial constraints. 

Our approach differs radically from traditional correlation and fluctuation analyses, which are not done at fixed $b$. 
In order to reduce the dependence on $b$, 
one usually forms combinations of multiplicities~\cite{Gorenstein:2011vq}, such as  ratios~\cite{Jeon:1999gr,Jeon:2000wg,ALICE:2017jsh} or linear combinations whose average value is zero, that go under the name of $\nu_{dyn}$~\cite{Pruneau:2002yf,STAR:2003oku,ALICE:2021fpb}. 
These procedures entail a huge loss of information.
From two observables, one typically extracts only one fluctuation measure, out of three that are relevant (the fluctuation of each observable, and their mutual correlation). 
 
The Bayesian reconstruction of impact parameter was introduced~\cite{Das:2017ned} in the context of ultrarelativistic nucleus-nucleus collisions, and it has also been implemented in collisions at lower energies~\cite{INDRA:2020kyj,Parfenov:2021ipw,Li:2022mni}.
It has then been extended  to proton-nucleus collisions~\cite{Rogly:2018ddx} and to correlation studies in nucleus-nucleus collisions~\cite{Yousefnia:2021cup}. 
Here we study correlations in proton-nucleus collisions, by extending the work of Ref.~\cite{Yousefnia:2021cup} to these collisions, along the lines of Ref.~\cite{Rogly:2018ddx}. 
 
The key ingredient of the reconstruction is to parametrize the fluctuations at fixed impact parameter in a way that is both simple and general.
We refer to this as to the fluctuation kernel.
A Gaussian kernel is good enough for nucleus-nucleus collisions~\cite{Das:2017ned,Yousefnia:2021cup}, but it must be replaced with a gamma kernel for proton-nucleus collisions~\cite{Rogly:2018ddx}.
In Sec.~\ref{s:kernel}, we introduce a simple generalization of the gamma kernel to several variables.
The accuracy of this parametrization is evaluated through simulations with the Angantyr model~\cite{Bierlich:2018xfw}.
In Sec.~\ref{s:bayesian}, we present the Bayesian method to reconstruct fluctuations and correlations in proton-nucleus collisions at $b=0$ from minimum-bias data. 
The method is validated using simulations with Angantyr, then applied to ATLAS data~\cite{ATLAS:2014qaj}, in which the two observables in each collision are the charged multiplicity in the central tracker, and the transverse energy in the forward calorimeter.\footnote{As we will argue in Sec.~\ref{s:relativecovariance}, the fluctuations of the transverse energy are likely to reflect those of the multiplicity in the corresponding pseudorapidity window.}
Results on multiplicity fluctuations and correlations at $b=0$ are presented in Sec.~\ref{s:models}. 
We compare ATLAS data with quantitative predictions from the Angantyr model. 
In order to link observations with the underlying QCD dynamics, we also show results from a simplified model that can be derived from high-energy QCD, and from the fluctuating-string model. 
We discuss how our results relate to previous analyses of long-range correlations. 
We argue that our picture provides a natural explanation for the observed centrality dependence of the multiplicity in p+Pb collisions. 

\section{Fluctuation kernel}
\label{s:kernel}

\begin{figure*}[ht]
\begin{center}
\includegraphics[width=.49\linewidth]{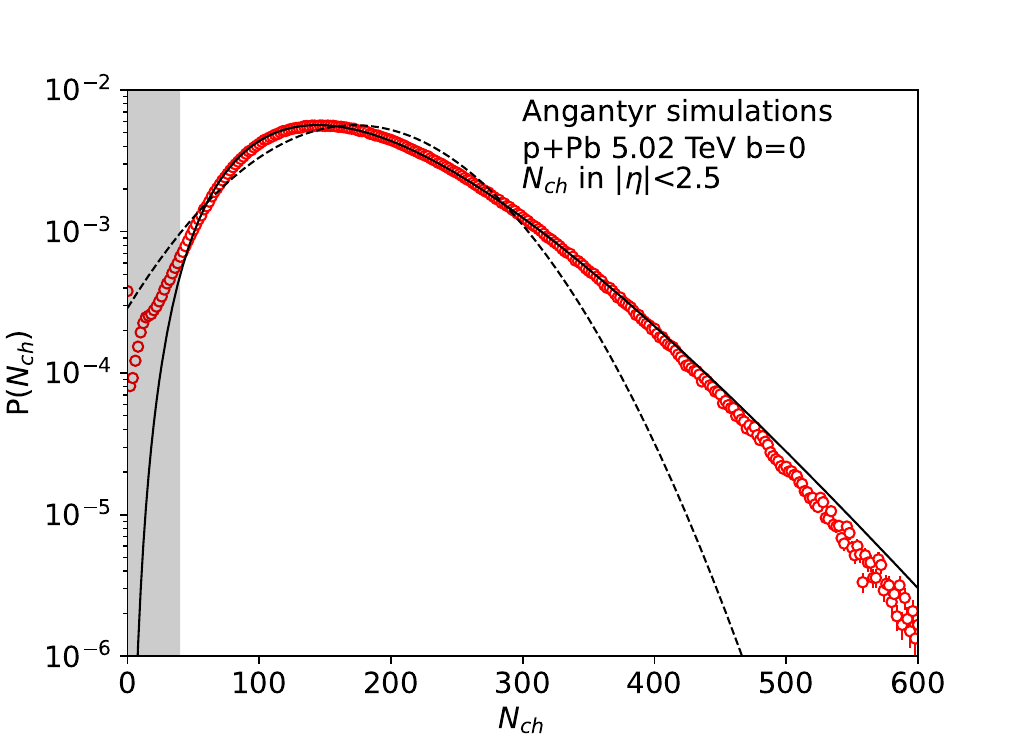}
\includegraphics[width=.49\linewidth]{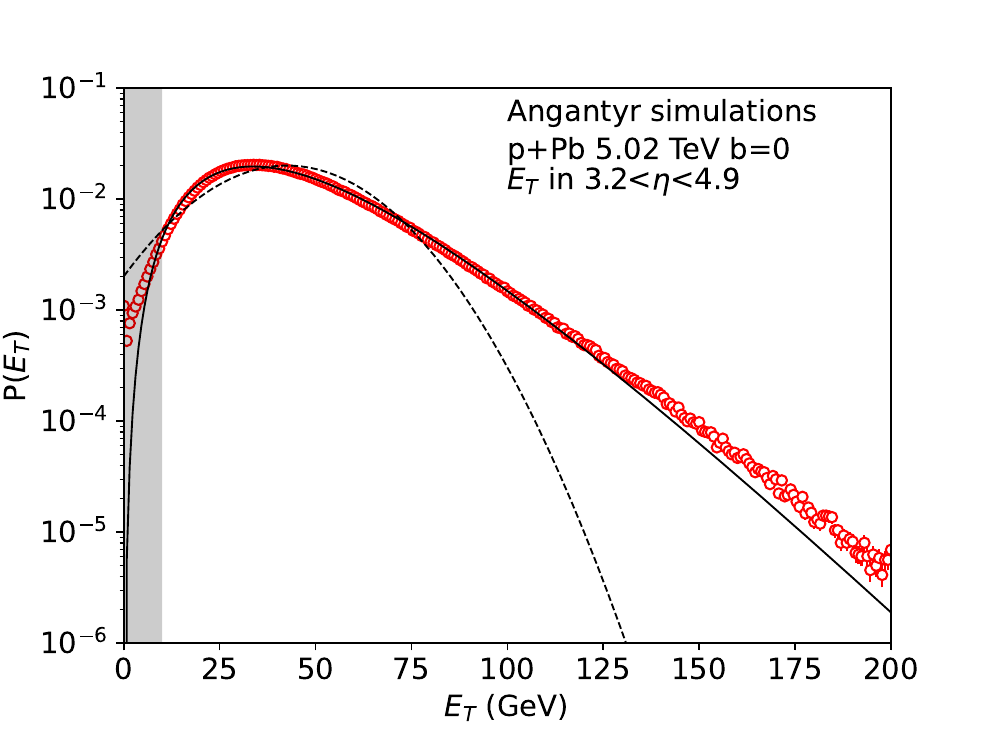}
\end{center}
\caption{(Color online)
  Probability distribution of the charged multiplicity $N_{ch}$ at central rapidity (left) and transverse energy $E_T^{\rm Pb}$ at forward rapidity (right) in p+Pb collisions at $\sqrt{s_{\rm NN}}=5.02$~TeV and $b=0$.
  Symbols: Angantyr simulation~\cite{Bierlich:2018xfw}.
  Lines:  two-parameters fits using gamma distributions (solid lines) or Gaussians (dotted lines).
  The shaded bands correspond to the range of values $N_{ch}$ and $E_T^{\rm Pb}$ that are excluded from the fit. 
}
\label{fig:projections}
\end{figure*}

\subsection{One variable}
\label{s:projected}

Fluctuations in a large system are Gaussian by virtue of the central limit theorem, so that the Gaussian is a natural choice for parametrizing the fluctuations of a random variable. 
The gamma distribution provides however a better parametrization for random variables that are positive, such as the multiplicity or transverse energy in a detector.
Like the Gaussian, the gamma distribution is fully determined by its mean and standard deviation, but unlike the Gaussian, it has positive support (see Appendix~\ref{s:mapping}).
Both distributions coincide when the standard deviation is much smaller than the mean.
This is typically the case for large systems, so that the central limit is automatically verified also with a gamma distribution. 

Alternative distributions with positive support have been used in the context of high-energy physics, most notably, the negative binomial distribution (NBD)~\cite{Giovannini:1985mz,Bozek:2013uha}.
The gamma distribution can be viewed as a continuous version of the NBD, as explained in detail in Ref.~\cite{Rogly:2018ddx}.
It is therefore more flexible than the NBD in the sense that it also applies to continuous variables, such as the transverse energy. 

In order to illustrate the usefulness of the gamma distribution, we have generated $6\times 10^6$ central ($b=0$) p+Pb collisions at $\sqrt{s_{\rm NN}}=5.02$~TeV using the Angantyr model~\cite{Bierlich:2018xfw}.\footnote{Note that the laboratory frame does not coincide with the center-of-mass frame of nucleon-nucleon collisions, because of the common magnetic field and different charge to mass ratios. In the laboratory frame, the energy of the proton is $E_p=\frac{1}{2}\sqrt{(A/Z)s_{\rm NN}}$ and the energy per nucleon of the Pb nucleus is $E_{\rm Pb}=\frac{1}{2}\sqrt{(Z/A)s_{\rm NN}}$, with $Z=82$ and $A=208$.}
Angantyr is the heavy-ion version of \textsc{Pythia}~8~\cite{Sjostrand:2014zea,Bierlich:2022pfr}, which is a state-of-the-art microscopic description of hadronic collisions.
For each collision event,\footnote{The calculations were based on the example {\tt main113.cc}, which is part of the official PYTHIA release, and the exact same parameters were used.} we compute two observables analogous to those measured by the ATLAS collaboration~\cite{ATLAS:2014qaj}: 
the charged multiplicity $N_{ch}$ in the pseudorapidity interval $|\eta|<2.5$, and the transverse energy (of charged and neutral particles) $E_T^{\rm Pb}$ in the pseudorapity interval $3.2<\eta<4.9$, where positive values of $\eta$ correspond to the Pb-going side.\footnote{We follow the convention of the ALICE Collaboration~\cite{ALICE:2014xsp}, which is opposite to that of ATLAS~\cite{ATLAS:2015hkr}.} 
The distributions of these two observables are displayed in Fig.~\ref{fig:projections}. 
Each distribution is fitted with a gamma distribution.
The fits are of good quality, except for the lowest values of $N_{ch}$ and $E_T^{\rm Pb}$. 
One usually eliminates these low values anyway when fitting experimental data (see Ref.~\cite{Yousefnia:2021cup} and Sec.~\ref{s:atlas} below), so that we exclude them from the fit.

\subsection{Quantifying the difference between fit and data}
\label{s:KL}

Throughout this paper, we carry out standard $\chi^2$ fits to Monte Carlo simulations or experimental data. 
However, we eventually want to quantify {\it systematic\/} deviations between fit and data (or simulations), and the $\chi^2$ is not appropriate since one divides the deviation by the  {\it statistical\/} error. 
An alternative measure of the difference between the probability distribution $P_{\rm data}$ given by the data (or the simulations), and the fit to this distribution $P_{\rm fit}$, is the Kullback-Leibler divergence~\cite{Kullback:1951zyt}
\begin{equation}
  \label{DKL}
D_{\rm KL}\equiv\sum_i P_{\rm data}(i)\ln\left(\frac{P_{\rm data}(i)}{P_{\rm fit}(i)}\right),
\end{equation}
where the sum runs over all bins.
Both probability distributions must be normalized:
\begin{equation}
\sum_i P_{\rm data}(i)=\sum_i P_{\rm fit}(i)=1.
\end{equation}
The Kullback-Leibler divergence has a simple interpretation when the relative difference between $P_{\rm data}(i)$ and $P_{\rm fit}(i)$ is small for all $i$.
Then, one can write $P_{\rm fit}(i)=P_{\rm data}(i)(1+\varepsilon(i))$, where $|\varepsilon(i)|\ll 1$ is the relative difference between fit and data.
Inserting into Eq.~(\ref{DKL}) and expanding to lowest non-trival order in $\varepsilon(i)$, one obtains:
\begin{equation}
\label{expansion}
  D_{\rm KL}\simeq\frac{1}{2}\sum_i P_{\rm data}(i)\varepsilon(i)^2=\frac{1}{2}\langle \varepsilon^2\rangle,
\end{equation}
where angular brackets denote the average over bins.
Throughout this article, we use the following quantity as a measure of the accuracy of the fit, and refer to it loosely as to the ``rms error'' because it corresponds to the rms value of $\varepsilon$ when the error is small (note that even when the error is small, it is in fact the rms {\it relative\/} error, that is, the rms error on a logarithmic plot):
\begin{equation}
  \label{rmserror}
  (2 D_{\rm KL})^{1/2},  
\end{equation}
where $D_{KL}$ is evaluated using Eq.~(\ref{DKL}). 
We always choose bins which are wide enough that the statistical fluctuations in each bin contribute little to this rms error. 

The rms errors of the fits presented in Fig.~\ref{fig:projections} are $3.1\%$  and $4.4\%$ for the distributions of $N_{\rm ch}$ and $E_T^{\rm Pb}$, respectively.
If we replace the gamma distribution with a Gaussian, these errors become $52.2\%$ and $36.5\%$, so that the gamma distribution increases the precision of the fits by an order of magnitude, at no additional cost in terms of fit parameters. 

The tail of the $E_T$ distribution is above the fit. 
This discrepancy can be put down to hard scatterings, which create particles with high $E_T$, and whose probability decreases like a power law at large $E_T$~\cite{Field:1976ve}, slower than the gamma distribution which is exponential.
On the other hand, the tail of the $N_{ch}$ distribution is below the fit, and it is tempting to postulate that this is again due to hard processes, which produce a large $E_T$, but few particles.

\begin{figure*}[ht]
\begin{center}
\includegraphics[width=.49\linewidth]{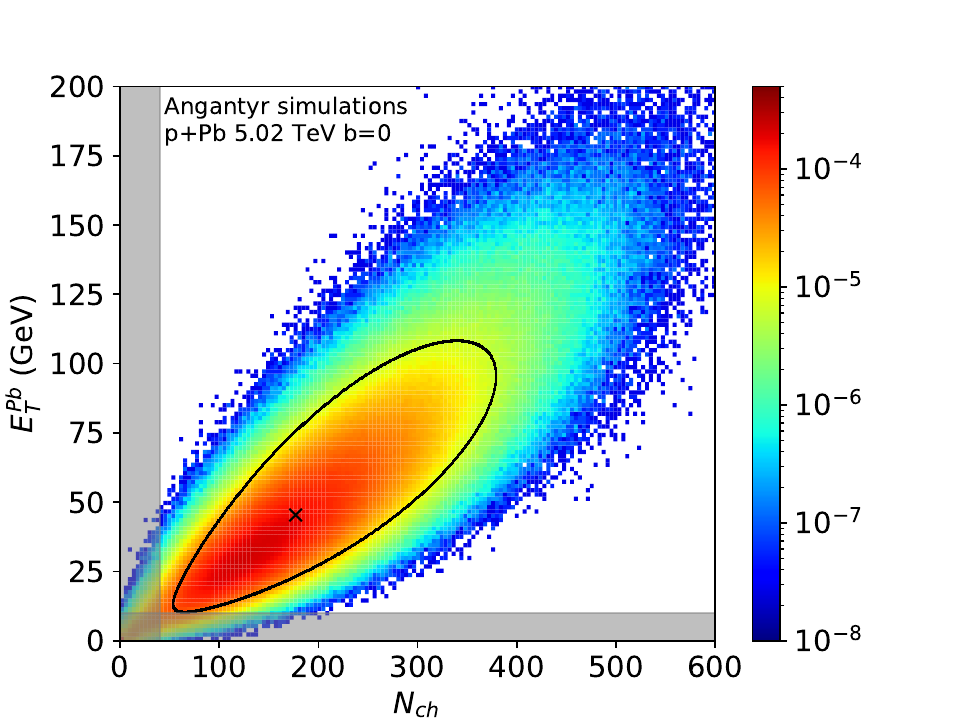}
\includegraphics[width=.49\linewidth]{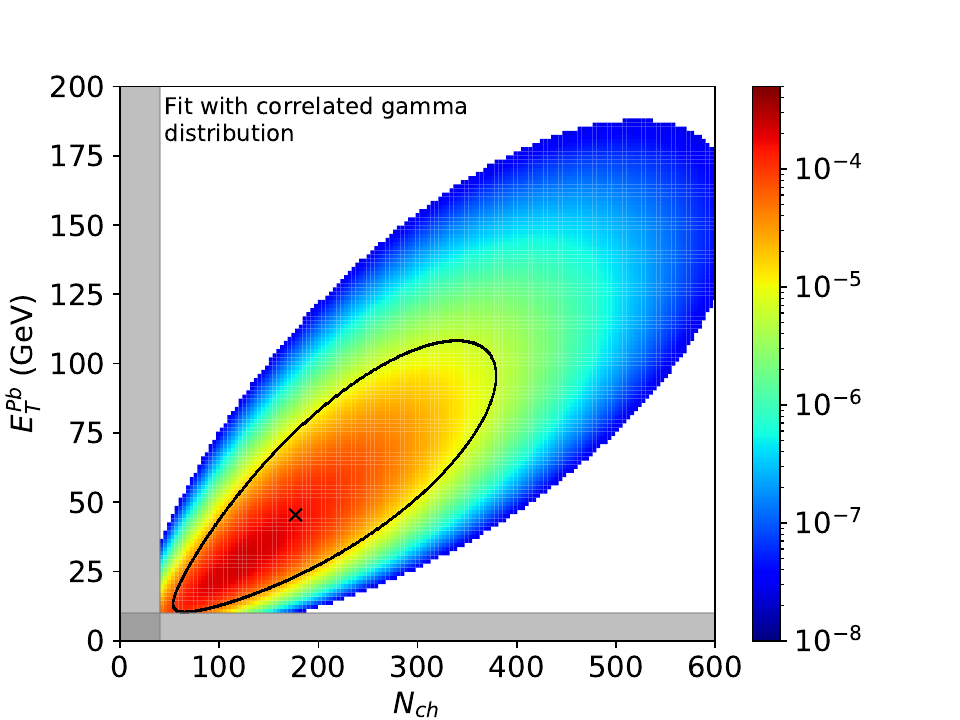}
\end{center}
\caption{(Color online)
Left: two-dimensional representation of the Angantyr simulation shown in Fig.~\ref{fig:projections}. 
The correlated gamma fit is indicated by the cross, which is the mean value, and by the black line around it, which encompasses the 90\% confidence area. 
We define this area by mapping the 90\% confidence
ellipse of the Gaussian distribution (\ref{multigaussian}) to the gamma variables according to Eq. (\ref{mapping}) (see
Appendix~\ref{s:confidence} for details). 
Right: the fitted correlated gamma distribution.
 As in Fig.~\ref{fig:projections}, the shaded area, corresponding to the lowest values of $N_{ch}$ and $E_T^{\rm Pb}$, is excluded from the fit.
  Since the fitting function is smooth, it does not present the statistical fluctuations that appear as scattered points on the left plot. 
  The color is set to white for bins in which the mean number of events is smaller than 1.  
  The same convention holds for similar plots below. 
  }
\label{fig:b=0pPb}
\end{figure*}

\subsection{Extension to several variables}
\label{s:multigamma}

This joint distribution of $N_{ch}$ and $E_T^{\rm Pb}$ is displayed in Fig.~\ref{fig:b=0pPb} (left) for our Angantyr simulation of central p+Pb collisions.  
These two observables are strongly correlated, even though the impact parameter is fixed. 
We explain how to construct a simple parametrization of this correlated distribution.

While the Gaussian distribution can be readily generalized to an arbitrary number of correlated variables~\cite{Yousefnia:2021cup}, there is no standard generalization of the gamma distribution to several variables. 
We construct a correlated gamma distribution in the following way.
For a given variable $N$ whose distribution $P_\gamma(N)$ is a gamma distribution, we map it onto a variable $N'$ whose distribution $P_G(N')$ is a Gaussian with the same mean and variance.
This mapping is defined by matching the two cumulative distributions
(see Appendix~\ref{s:mapping} for details):
\begin{equation}
  \label{mapping}
  \int_{0}^NP_\gamma(x)dx=\int_{-\infty}^{N'}P_G(x)dx. 
\end{equation}
Once the mean and variance are specified, this equation defines $N'$ as a function of $N$.
Note that the lower bound of the integral is $0$ for the variable $N$, which has positive support, and $-\infty$ for the Gaussian variable $N'$, whose support is the real axis. 
Differentiating Eq.~(\ref{mapping}), one obtains 
\begin{equation}
\label{probacons1}
P_\gamma(N)dN=P_G(N')dN',
\end{equation}
which expresses the equality of elementary probabilities. 

In order to construct a correlated version of the gamma distribution, we start with a correlated Gaussian distribution for $p$ variables, $P_G(N'_1,\cdots,N'_p)$~\cite{Yousefnia:2021cup}: 
\begin{equation}
\label{multigaussian}
P_G(N'_1,...,N'_p)=
\frac{\exp\left(-\frac{1}{2}(N'_i-\bar N'_i)\Sigma^{-1}_{ij}(N'_j-\bar N'_j)\right)}
{\left((2\pi)^{p} |\Sigma| \right)^{1/2}},
\end{equation}
where, in the exponential, we use the Einstein summation convention over the repeated indices $i$ and $j$. 
In this equation, $\bar N'_i$ is the mean value of $N'_i$, and $\Sigma_{ij}$ is the symmetric covariance matrix. 
$\Sigma^{-1}$ denotes the inverse matrix and $\left|\Sigma\right|$ the determinant. 

The marginal distributions of a multivariate Gaussian distribution are themselves Gaussian, so that each of the variables $N'_1,\cdots,N'_p$ follows a Gaussian distribution: 
\begin{equation}
\label{singlegaussian}
P_G(N'_i)=\frac{1}{(2\pi\Sigma_{ii})^{1/2}}
\exp\left(-\frac{1}{2\Sigma_{ii}}(N'_i-\bar N'_i)^2\right).
\end{equation}
We then map each variable $N'_i$ onto a variable $N_i$ according to Eq.~(\ref{mapping}).

The correlated gamma distribution $P_\gamma(N_1,...,N_p)$ is finally defined by matching the elementary probabilities, through a straightforward generalization of Eq.~(\ref{probacons1}): 
\begin{equation}
  \label{probacons}
  P_G(N'_1,...,N'_p)dN'_1...dN'_p=P_\gamma(N_1,...,N_p)dN_1...dN_p.
\end{equation}
Using Eq.~(\ref{probacons1}) for each of the variables, we finally obtain: 
\begin{equation}
  \label{defpgamma}
P_\gamma(N_1,...,N_p)\equiv \frac{P_G(N'_1,...,N'_p)}{P_G(N'_1)...P_G(N'_p)}P_\gamma(N_1)...P_\gamma(N_p). 
\end{equation}
Note that the mean and variance of $N_i$ coincide with those of $N'_i$ by construction.
On the other hand, the covariance $\Sigma_{ij}$ in Eq.~(\ref{multigaussian}) does not coincide with the covariance of the correlated gamma distribution, which is somewhat smaller. 

\begin{figure}[ht]
\begin{center}
\includegraphics[width=\linewidth]{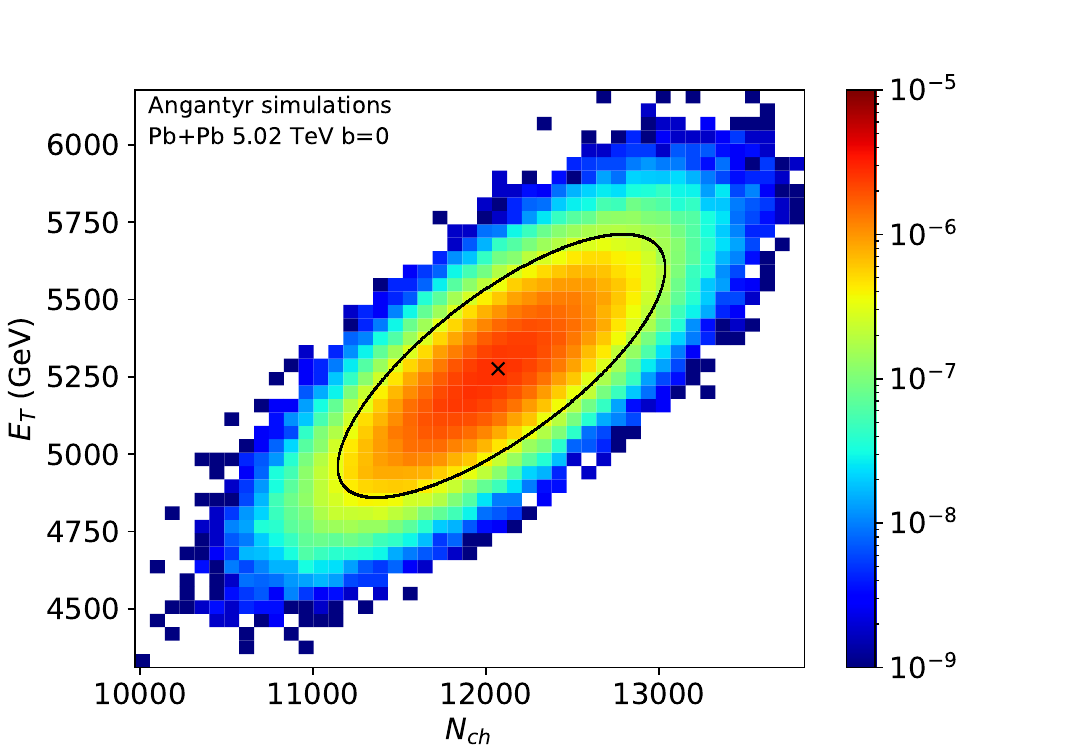}
\end{center}
\caption{(Color online)
  Distribution of $N_{ch}$ and $E_T$ for $3\times 10^5$ Pb+Pb collisions at $\sqrt{s_{\rm NN}}=5.02$~TeV and $b=0$ simulated with Angantyr.
  Here, $E_T$ denotes the sums of transverse energies deposited in
  $3.2<\eta<4.9$ and
  $-4.9<\eta<-3.2$,
  corresponding to the forward and backward calorimeters of ATLAS~\cite{ATLAS:2019peb}.
 As in Fig.~\ref{fig:b=0pPb}, the black line represents the 90\% confidence area of the correlated gamma fit (not shown). 
  }
\label{fig:b=0PbPb}
\end{figure}

The correlated gamma distribution can be defined for an arbitrary number $p$ of random variables, but in the remainder of this paper, we only implement the simplest case $p=2$, where $N_1\equiv N_{ch}$  and $N_2\equiv E_T^{\rm Pb}$. 
There are five parameters: the means and variances of each variable, and their mutual correlation, which is encoded in the off-diagonal element $\Sigma_{12}$ in Eq.~(\ref{multigaussian}). 
The fit to Angantyr results with a correlated gamma distribution is displayed in the right panel of Fig.~\ref{fig:b=0pPb}. 
It is obtained by carrying out a five-parameter fit to the distribution on the left. 
However, four out of these five parameters are already known from the marginal distributions in Fig.~\ref{fig:projections}, so that the only extra fit parameter is  $\Sigma_{12}$. 
One sees by eye that the correlated gamma distribution captures the main features of the simulation.  
A closer examination reveals that the fit is not perfect.
As already seen in Fig.~\ref{fig:projections}, the fit has a longer tail in the $N_{ch}$ direction, and a shorter tail in the $E_T$ direction. 
The rms error is $12.3\%$. It is significantly larger than for the projections, but much smaller than the rms error of a two-dimensional Gaussian fit (not shown) which is $46.0\%$. 
Thus, the quality of the fit is improved by a factor $\sim 4$ by replacing the correlated Gaussian distribution with a correlated gamma distribution, at no extra cost.

For the sake of comparison, we have also carried out a simulation of Pb+Pb collisions at $b=0$. 
The joint distribution of $N_{ch}$ and $E_T$ for this simulation is displayed in Fig.~\ref{fig:b=0PbPb}. 
The system size is much larger, as can be seen by comparing the values of $N_{ch}$ and $E_T$ with those in Fig.~\ref{fig:b=0pPb}.
Correspondingly, the relative fluctuations are smaller and more Gaussian~\cite{Yousefnia:2021cup}. 
We have carried out fits (not shown) with a correlated Gaussian and with a correlated gamma distribution. 
The rms errors are $6.3\%$ and $5.9\%$. 
The small difference between these two errors means that for this system, the correlated gamma distribution is a marginal improvement over the Gaussian distribution. 
This is not surprising since both distributions coincide for a large system. 
The error with the correlated gamma distribution is smaller by a factor $\sim 2$ than for the p+Pb collisions, which is also not surprising as fluctuations in larger systems are more generic.  
 
Thus the correlated gamma distribution provides a decent parametrization of fluctuations at $b=0$, both for proton-nucleus and nucleus-nucleus collisions.
We have not run simulations for other values of $b$,\footnote{The currently available version of Angantyr does not allow one to run at fixed positive impact parameter.} but we expect that the correlated gamma parametrization still works, nevertheless getting worse as $b$ increases, similar to what is observed for Gaussian parametrizations in Pb+Pb collisions~\cite{Yousefnia:2021cup}. 

\begin{figure*}[ht]
\begin{center}
\includegraphics[width=.49\linewidth]{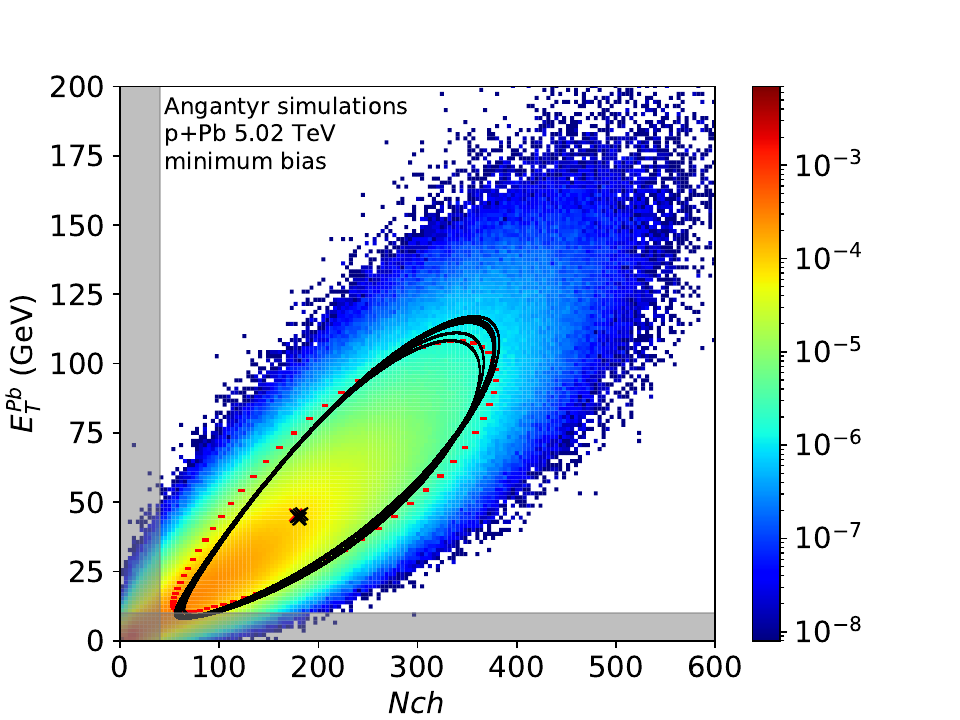}
\includegraphics[width=.49\linewidth]{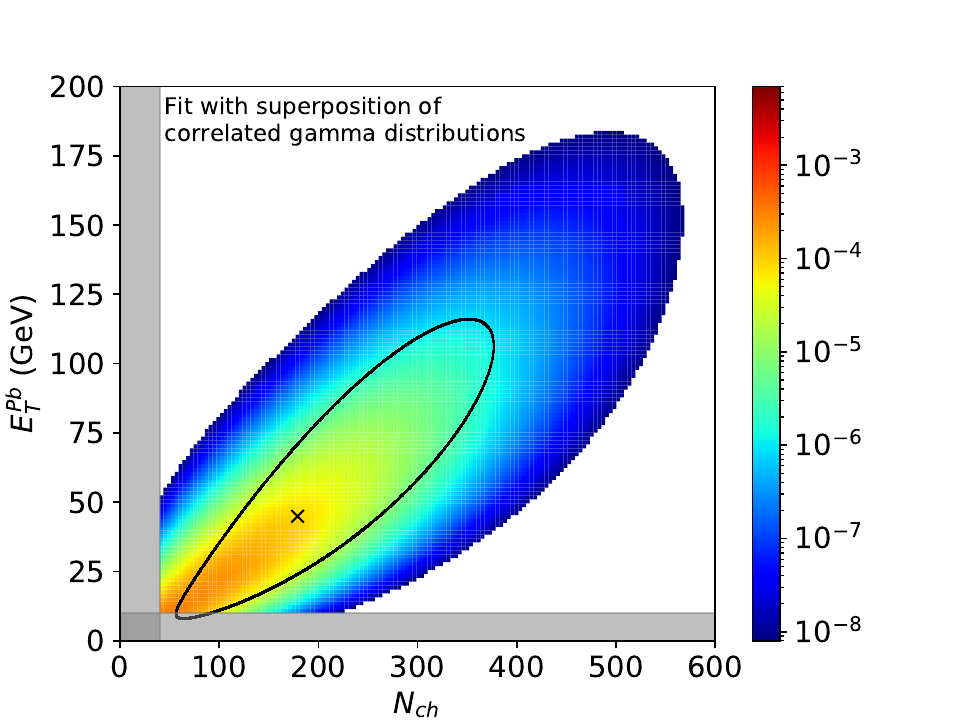}
\end{center}
\caption{(Color online)
  Joint probability distribution of $N_{ch}$ and $E_T^{\rm Pb}$ in minimum-bias p+Pb collisions at $\sqrt{s_{\rm NN}}=5.02$~TeV.
  Left panel: Angantyr simulations.
  Lines encompass the 90\% confidence area of $b=0$ collisions, either calculated directly (dotted line, corresponding to the solid line of Fig.~\ref{fig:b=0pPb}), or reconstructed (solid lines, where each line corresponds to a different initialization of the fit parameters, corresponding to the range of parameters in Table~\ref{fitparam}).
  The right panel displays the fit obtained using Eq.~(\ref{minbias}) with one specific initialization, and the corresponding 90\% confidence area of $b=0$ collisions. 
      }
\label{fig:MBpPb}
\end{figure*}

\section{Bayesian reconstruction of $b=0$ collisions}
\label{s:bayesian}

We now describe how correlations and fluctuations at $b=0$ can be inferred from minimum-bias data.
First, we explain why we choose the impact parameter as a centrality criterion, rather than, say, the number of participant nucleons, which may be thought a more relevant measure of the collision activity~\cite{Miller:2007ri}.
The reason is that:
\begin{itemize}
\item Any observable defined from an ensemble average at fixed $b$ (for instance, the average value, or the variance, of the multiplicity in a given rapidity interval) is a smooth function of $b^2$ due to $b\to -b$ symmetry.
\item The cumulative distribution of $b$, which we denote by $c_b$~\cite{Das:2017ned}, is itself proportional to $b^2$ near $b=0$:
  \begin{equation}
    \label{cb}
    c_b=\frac{\pi b^2}{\sigma_{\rm inel}}, 
    \end{equation}
  where $\sigma_{\rm inel}$ denotes the total inelastic cross section of the collision.\footnote{In this paper, we focus on results for $b=0$, which are independent of the value of $\sigma_{\rm inel}$.}
\end{itemize}
Therefore, any ensemble-averaged observable is a smooth function of $c_b$ and has no singularity at $c_b=0$. 
This simple and robust property, which is the only input of our reconstruction, is not satisfied with alternative definitions of the centrality.

\subsection{Method}
\label{s:method}

We assume that for a fixed value of $b$ or, equivalently, $c_b$, the probability distribution of the observables $N_i$ of interest (in our case, $N_{ch}$ and $E_T^{\rm Pb}$) is a correlated gamma distribution as constructed in Sec.~\ref{s:multigamma}, which we denote by $P_{\gamma}(N_1,N_2|c_b)$.
This correlated gamma distribution has five parameters, which are the five parameters of the Gaussian distribution (\ref{multigaussian}), and which we denote collectively by $\Pi_j(c_b)$, $j=1,..., 5$.
We assume that each of these parameters is a smooth function of $c_b$, without any singularity at $c_b=0$. 
We parametrize these functions in a way that is as general as possible. 
The parametrization we choose is  the exponential of a polynomial~\cite{Yousefnia:2021cup}, which guarantees that $\Pi_j(c_b)>0$. 
The degree of the polynomial must be large enough in order to obtain a satisfactory fit to the data. 
Here, a polynomial of degree 2 was found to be sufficient: 
\begin{equation}
\label{cbdependence}
  \Pi_j(c_b)=\Pi_j(0)\exp\left(-a_{1,j} c_b-a_{2,j} c_b^2\right).
\end{equation}
The probability distribution of $N_1$ and $N_2$ in minimum-bias collisions is then obtained by integrating over $c_b$:
\begin{equation}
\label{minbias}
P(N_1,N_2)=\int_0^1 P_{\gamma}(N_1,N_2|c_b)dc_b. 
\end{equation}
We fit the left-hand side, as obtained from simulation or experimental data, with the right-hand side.
According to Eq.~(\ref{cbdependence}), there are 3 fit parameters for each of the parameters of the gamma distribution, so that there is a total of 15 fit parameters.
The only constraint that we impose on the fit parameters is that the mean values of $N_1$ and $N_2$ decrease with increasing impact parameter.

\subsection{Validation using Angantyr simulations}
\label{s:validation}

We now assess the accuracy of the Bayesian reconstruction by applying it to $2\times 10^7$ minimum-bias p+Pb collisions at $\sqrt{s_{\rm NN}}=5.02$~TeV simulated with Angantyr. 
Each quantity reconstructed at $b=0$ can be compared with that calculated directly by simulating events at $b=0$, so that we can assess quantitatively the accuracy of the reconstruction. 

The distribution of $N_{ch}$ and $E_T^{\rm Pb}$ in minimum-bias collisions is displayed in the left panel of Fig.~\ref{fig:MBpPb}. 
At first sight, it looks similar to the distribution in collisions at $b=0$, shown in Fig.~\ref{fig:b=0pPb}. 
Therefore, it is not surprising that some information about $b=0$ collisions can be reconstructed from minimum-bias events. 

\begin{table}[ht]
\begin{tabular}{|c||c|c|c|}
\hline
& reconstruction
&direct calculation
&max. error\cr
\hline
\hline
$\bar N_{ch}$
& $178.0-183.1$
&  $177.0$
&$3.4\%$ \cr 
\hline
$\sigma_{N_{ch}}$
& $70.1-76.1$
& $76.8$
& $9.5\%$ \cr 
\hline
$\bar E_T^{\rm Pb}$
& $43.9-45.8$~GeV
& $45.4$~GeV
&$3.3\%$\cr 
\hline
$\sigma_{E_T}$
& $23.5-25.8$~GeV
& $23.2$~GeV
&$11.2\%$\cr 
\hline
Pearson $r$
& $0.844-0.865$
& $0.846$
&$2.2\%$\cr  
\hline
\end{tabular}
\caption{\label{fitparam}
  Comparison between values reconstructed from simulations of minimum-bias collisions (Fig.~\ref{fig:MBpPb}), or calculated directly by simulating p+Pb collisions at $b=0$ (see Fig.~\ref{fig:b=0pPb}),  for several observables: average values and standard deviations of $N_{ch}$ and $E_T^{\rm Pb}$, Pearson correlation coefficient $r\equiv\Sigma_{12}/(\sigma_{E_T}\sigma_{N_{ch}})$. 
  For reconstructed values, we obtain a range depending on the initialization of the fit parameters.
  The last column is the maximum error on each observable.
}
\end{table}

We carry out a 15-parameter fit of this minimum-bias distribution, as explained in Sec.~\ref{s:method}.
The number of fit parameters is too large for their values to be precisely constrained by simulated data.
In practice, the set of fit parameters returned by the fitting algorithm depends somewhat on the initial guess, which means that the algorithm does not find the absolute minimum of the $\chi^2$.
We have not attempted to solve this numerical issue since the fits, despite not being identical, are all of equivalent quality, and the quality is excellent. 
Specifically, the rms error of the fit lies between $4.2\%$ and $7.3\%$, significantly smaller than when fitting $b=0$ simulations alone in Fig.~\ref{fig:b=0pPb}.
The 10 additional parameters introduced to model the $b$ dependence partially compensate for the imperfection of the correlated gamma distribution. 

We are eventually interested in $b=0$ collisions, which are described by only 5 out of the 15 fit parameters ($\Pi_j(0)$ in Eq.~(\ref{cbdependence})). 
Table~\ref{fitparam} lists the ranges of values for these parameters for the various initial guesses, as well as the reference value obtained by simulating events at $b=0$. 
The maximum error in the right column is defined as the largest relative difference between the reconstructed value and the reference value. 
The average values of $N_{ch}$ and $E_T^{\rm Pb}$ at $b=0$ are reconstructed with $\sim 3\%$ accuracy, and their standard deviations with $\sim 10\%$ 
accuracy.\footnote{The error on the reconstruction of the mean value was larger by a factor $\sim 3$ in Ref.~\cite{Rogly:2018ddx}. 
  The reason is twofold.
  First, it was assumed in Ref.~\cite{Rogly:2018ddx} that the ratio of variance to mean was independent of centrality.
  Here, we relax this assumption, following Ref.~\cite{Yousefnia:2021cup}. 
  The variances and the covariance then decrease with $b$ faster than the mean, which is probably not physical~\cite{Yousefnia:2021cup} but results in a much improved reconstruction of $b=0$ quantities, which is our goal. 
  Second, the reconstruction of mean values and variances is more precise if one fits the joint distribution of $(N_{ch},E_T^{\rm Pb}$) than if one fits the marginal distributions. 
} 
It is quite remarkable that these quantities can be reconstructed so well with such minimal input. 

\begin{figure*}[ht]
\begin{center}
\includegraphics[width=.49\linewidth]{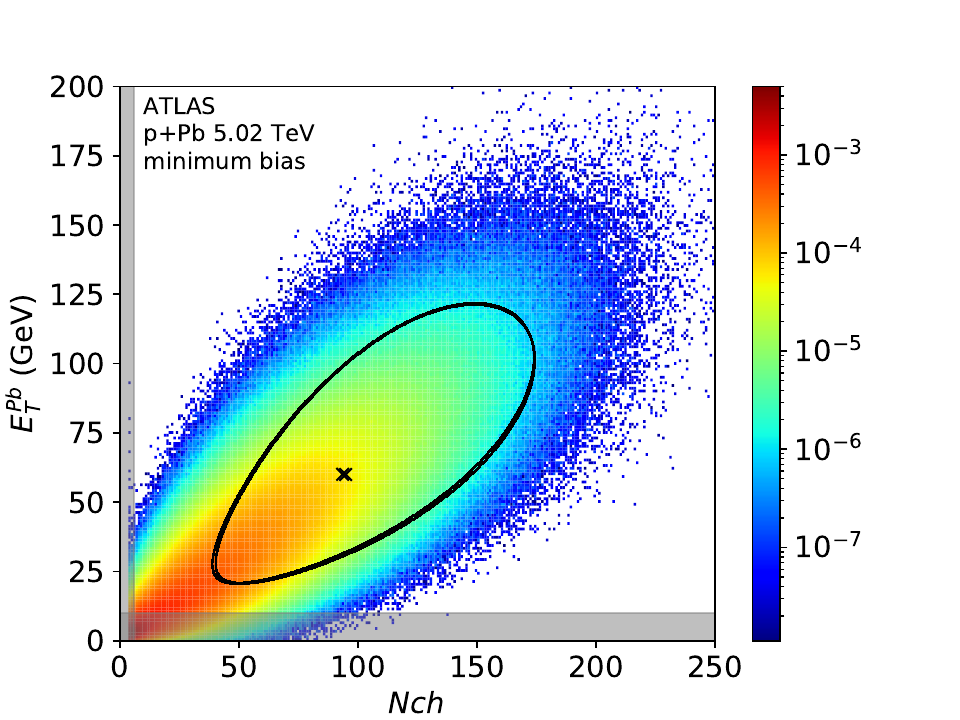}
\includegraphics[width=.49\linewidth]{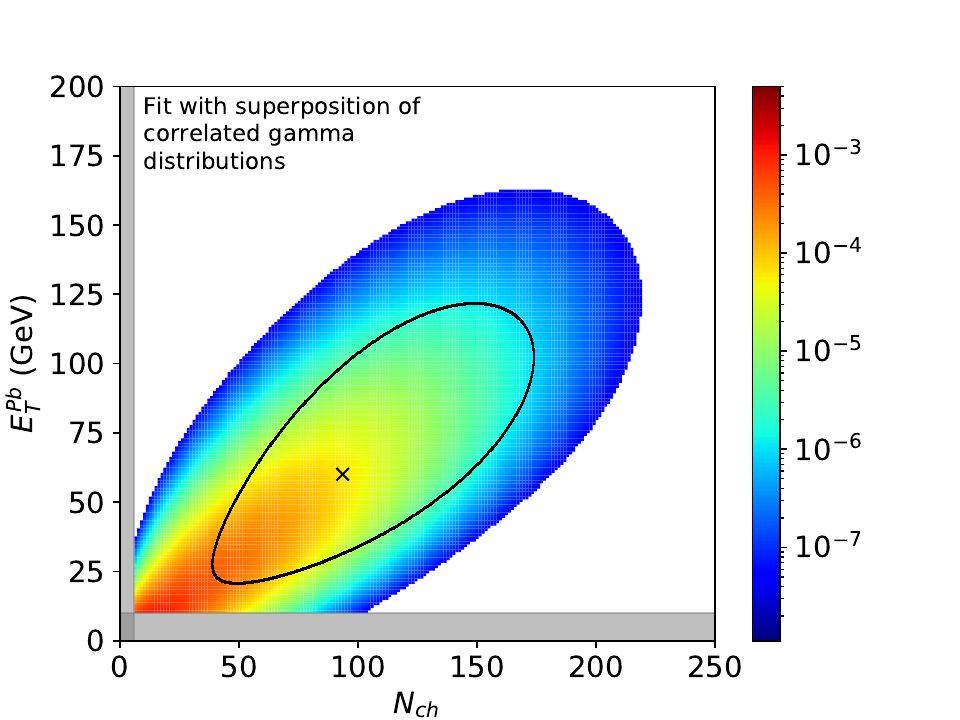}
\end{center}
\caption{(Color online)
  Joint probability distribution of $N_{ch}$ and $E_T^{\rm Pb}$ in minimum-bias p+Pb collisions at $\sqrt{s_{\rm NN}}=5.02$~TeV.
  Left: ATLAS data \cite{ATLAS:2014qaj}. 
  Two lines are drawn, which almost coincide, and correspond to the 90\% area of $b=0$ collisions given by the Bayesian reconstruction for two different initializations of the fit, as in Fig.~\ref{fig:MBpPb}.
  Right: fit using Eq.~(\ref{minbias}) for one specific initialization. 
We have applied the cuts $N_{ch}\ge 6$ and $E_T^{\rm Pb}>10$~GeV, 
and the excluded range is shown as a shaded area. 
In the right panel, the cut between the colored region and the white region has been placed arbitrarily (due to the various triggers used by ATLAS, the rule of one event per bin used in previous plots was ambiguous). 
  }
\label{fig:ATLASpPb}
\end{figure*}

\subsection{Application to ATLAS data}
\label{s:atlas}

The left panel of Fig.~\ref{fig:ATLASpPb} displays the distribution of $N_{ch}$ and $E_T^{\rm Pb}$ measured by ATLAS in minimum-bias p+Pb collisions at $\sqrt{s_{\rm NN}}=5.02$~TeV~\cite{ATLAS:2014qaj}. 
The central detector of ATLAS only detects a fraction of the charged particles in the interval $|\eta|<2.5$, which explains why values of $N_{ch}$ are much lower than in the Angantyr simulation in Fig.~\ref{fig:MBpPb}. 
The fit using Eq.~(\ref{minbias}) is displayed in the right panel.
Compared to Angantyr simulations, the sensitivity of final results to initial guesses is much reduced.
We attribute this to the fact that fluctuations are smaller in ATLAS data than in Angantyr simulations, a point to which we come back in Sec.~\ref{s:models}. 
Reconstructed mean values at $b=0$ are $\bar N_{ch}=94$, $\bar E_T^{\rm Pb}=60$~GeV. Standard deviations are $\sigma_{N_{ch}}=32$ and $\sigma_{E_T}=24$~GeV, and the Pearson correlation coefficient between $N_{ch}$ and $E_T^{\rm Pb}$ is $r=0.65$.
The rms error of the fit is 4.5\%.
Based on the validation in Sec.~\ref{s:validation}, we expect that the accuracy of the reconstruction is of order $3\%$ for mean values, and $10\%$ for standard deviations. 

We finally explain why we only reconstruct observables at zero impact parameter, not their full impact parameter dependence.
First, it is impossible to reconstruct the impact parameter dependence of the covariance matrix from minimum-bias data alone. 
The reason is that for $b>0$, a simultaneous increase in $N_{ch}$ and $E_T$ can be produced either by an decrease of $b$, or by a fluctuation at fixed $b$, and both effects cannot be disentangled.
In the case of nucleus-nucleus collisions, a detailed study~\cite{Yousefnia:2021cup} has shown that the impact parameter dependence can be reconstructed only for the mean values, and for a specific projection of the covariance matrix representing the width of the distribution of $(N_{ch},E_T)$.
We have not attempted to extend this study to proton-nucleus collisions,\footnote{One reason is that the Angantyr model which we use for the validation cannot be run at fixed $b$ for $b>0$.} but we expect that conclusions would be qualitatively similar.
This suggests that the parameters $a_{1,j}$ and $a_{2,j}$ in Eq.~(\ref{cbdependence}) are of little significance if $j$ labels an element of the covariance matrix, but may contain relevant information for the mean values $\bar N_{ch}$ and $\bar E_T$.
We obtain $a_1\simeq 1.6$ and $a_2\simeq 0.7$ for $\bar N_{ch}$, $a_1\simeq 1.0$ and $a_2\simeq 1.8$ for $\bar E_T$. 
To leading order, the dependence on impact parameter is determined by $a_1$. 
We find that $a_1$ is smaller for $E_T$, in the Pb-going rapidity region, than for $N_{ch}$, around central large rapidity. 
This suggests that the dependence on impact parameter becomes weaker as one gets closer to the rapidity of the Pb nucleus. 

\section{Results and model comparisons} 
\label{s:models}

\subsection{Relative fluctuations in central p+Pb collisions} 
\label{s:relativecovariance}

The observable which we use to characterize fluctuations and correlations is the relative covariance matrix $\sigma_{ij}$, also referred to as the robust covariance~\cite{Pruneau:2002yf}. 
If $N_1$ and $N_2$ denote multiplicities in two separate rapidity intervals, it is defined by 
\begin{equation}
  \label{relcovariance}
  \sigma_{ij}=\frac{\langle N_iN_j\rangle-\langle N_i\rangle\langle N_j\rangle-\langle N_i\rangle\delta_{ij}
  }{\langle N_i\rangle\langle N_j\rangle},
\end{equation}
where $i,j=1,2$, and angular brackets denote an ensemble average, in our case an average over $b=0$ collisions.
The last term in the numerator subtracts self-correlations from the variance~\cite{Pruneau:2002yf}, and isolates the dynamical fluctuations. 
After this subtraction is carried out, the relative variance depends little on the size of the rapidity bin (see Fig.~\ref{fig:dipole} for an illustration). 
It is also independent of the detector efficiency, which is a constant factor canceling between the numerator and the denominator~\cite{Pruneau:2002yf}.
Therefore, it makes sense to compare ATLAS data directly with Angantyr simulations, even if the ATLAS detector only detects a fraction of the charged particles. 


Eq.~(\ref{relcovariance}) can be readily generalized to observables other than multiplicities, such as the transverse energy $E_T$ in a calorimeter, which is obtained by summing the contributions of all particles falling in the acceptance window: $E_T=\sum_i E_{T,i}$, where $E_{T,i}$ is the transverse energy of particle $i$. 
The only modification lies in the self correlation, i.e., the last term of the numerator of Eq.~(\ref{relcovariance}), for which one must substitute~\cite{Yousefnia:2021cup,Cody:2021cja}:
\begin{equation}
  \label{selfcorrgeneral}
\langle N_i\rangle \rightarrow \left\langle\sum_i (E_{T,i})^2\right\rangle. 
\end{equation}
A calorimeter measures the energy without resolving the contributions of individual particles, therefore the right-hand side cannot be measured.
It can however be estimated.
In the case of ATLAS p+Pb data, we take it from the Angantyr calculation.
It is a small relative correction (of order $6\%$) to the variance of $E_T^{\rm Pb}$.

\newcommand\illustration[4]{
  \begin{tikzpicture}
    \def\proton{(0,0) ellipse [x radius=#1*#4,y radius=#2*#4]}
    \draw \proton;
    \fill [gray!20] \proton;
    \pattern [pattern=horizontal lines] \proton;
    \draw (0,0.4+#2*#4) node {#3 p};
    \def\nucleus{(4*#4,0) ellipse  [x radius=0.8*#4, y radius=3.2*#4]}
    \draw \nucleus;
    \fill [gray!50] \nucleus;
    \draw (4*#4,0.4+3.2*#4) node {Pb};
    \begin{scope}
      \clip (4.8*#4,-#2*#4) rectangle (1.6*#4,#2*#4);
      \fill [red!70] \nucleus;
      \pattern [pattern=horizontal lines, pattern color=black] \nucleus;
    \end{scope}
    \draw [dotted] (0,#2*#4) -- (4.8*#4,#2*#4);
    \draw [dotted] (0,-#2*#4) -- (4.8*#4,-#2*#4);
  \end{tikzpicture}
}

\begin{figure}
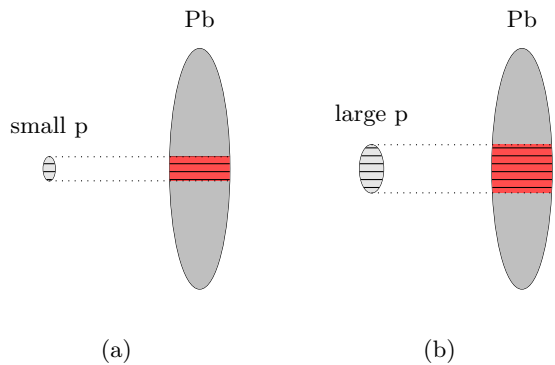

  \centering
  \begin{tabular}{ccc}
    \illustration{0.16}{0.32}{small}{.5}& \hspace{1cm} &\illustration{0.32}{0.64}{large}{.5}\\
    \\
    (a) & & (b)
  \end{tabular}
    \caption{(Color online)
Schematic representation of a p+Pb collision at $b=0$, where the proton size can be small (a) or large (b) depending on the event. 
Depending on the proton size, the number of wounded nucleons in the nucleus varies, as indicated by the 
relative volume of the interaction regions, colored in red.
  }
\label{fig:illustration}
\end{figure}

The relative covariance for collisions at $b=0$, reconstructed from ATLAS p+Pb data, is:
\begin{equation}
\label{atlascov}
  \sigma_{\rm ATLAS}=\left(
  \begin{matrix}
    0.101 & 0.084\cr
    0.084 & 0.147
  \end{matrix}\right)
\end{equation}
where the first variable is $N_{ch}$ and the second is $E_T^{\rm Pb}$.
We evaluate the maximum error on these figures to be $\sim 25\%$ by carrying out the same analysis as in Sec.~\ref{s:validation}.
This experimental result can be compared with that of Angantyr simulations of $b=0$ collisions: 
\begin{equation}
\label{angantyrcov}
  \sigma_{\rm Angantyr}=\left(
  \begin{matrix}
    0.182 & 0.187\cr
    0.187 & 0.244
  \end{matrix}\right).
\end{equation}
Angantyr overpredicts the relative (co)variances. 
In order to interpret this finding, let us list the physical mechanisms contributing to fluctuations in event generators.  
The number of nucleons hit by the incoming protons plays an essential role.  
In a Glauber calculation with a fixed nucleon-nucleon cross section~\cite{Miller:2007ri}, the nuclear volume traversed by the proton does not fluctuate, but the number of nucleons in this volume does,  corresponding to a quantum fluctuation associated with the nuclear wave function. 
If the cross section fluctuates event to event, the collision volume itself fluctuates, which entails much larger fluctuations in the number of wounded nucleons~\cite{Alvioli:2013vk}, as illustrated in Fig.~\ref{fig:illustration}. 
In the Angantyr model, these cross-section fluctuations are present, and correspond to fluctuations in the proton size.
Our study therefore suggests that these proton-size fluctuations are too large in Angantyr.\footnote{Note that contrary to a naive Glauber calculation, not all wounded nucleons are equivalent in Angantyr. 
The scatterings are softer in events where the proton is fatter~\cite{Bierlich:2016smv}, because the average number of Multi-Parton Interactions depends on the nucleon-nucleon impact parameter.}
This shows that proton-nucleus data constrain the theoretical description of subnucleonic fluctuations. 
It would be interesting to test other models of proton-nucleus collisions, such as the EPOS model~\cite{Pierog:2013ria}, or the hydrodynamic model with initial conditions from the 3D Glauber model~\cite{Shen:2017bsr,Shen:2022oyg}, in this context. 
 
One also observes that the relative variance is larger for $E_T$ than for $N_{ch}$, that is, $\sigma_{22}>\sigma_{11}$, both for data and Angantyr simulations.
This could be due to the fact that an energy ($E_T$) fluctuates more than a multiplicity ($N_{ch}$). 
In order to rule out this possibility, we have carried out Angantyr simulations at $b=0$ and calculated $E_T$ and $N_{ch}$ in the same pseudorapidity window. 
We have found that their relative variances are almost identical.
They differ only by $0.5\%$ in the window $|\eta|<2.5$, and by $3\%$ in the window $3.2<\eta<4.9$. 
We conclude that in a given pseudorapidity window, the relative variance of $E_T$ and the relative variance of $N_{ch}$ are equivalent observables. 
This should eventually be checked experimentally, by comparing the relative fluctuations of $N_{ch}$ and $\sum p_T$ in the same detector, as suggested in~\cite{Yousefnia:2021cup}.

The conclusion is that even though ATLAS does not measure the charged multiplicity in the window $3.2<\eta<4.9$, but only the transverse energy, the relative variance of the multiplicity would be almost identical, if measured.
The observation that $\sigma_{22}>\sigma_{11}$ therefore implies that the relative multiplicity fluctuations are larger at forward rapidity, on the Pb-going side, than in the central rapidity region. 
Now, the average multiplicity per unit rapidity $dN/dy$ is also larger on the Pb-going side in p+Pb collisions~\cite{ALICE:2014xsp}. 
One might naively think that the larger the multiplicity, the smaller its relative fluctuations. 
This reasoning works for statistical fluctuations, but we observe here the opposite trend for dynamical fluctuations.\footnote{ 
  A similar phenomenon was seen in $b=0$ Pb+Pb collisions~\cite{Yousefnia:2021cup}, where both the mean and the relative variance of the multiplicity are maximum at mid-rapidity.}

\begin{figure}[ht]
\begin{center}
\includegraphics[width=\linewidth]{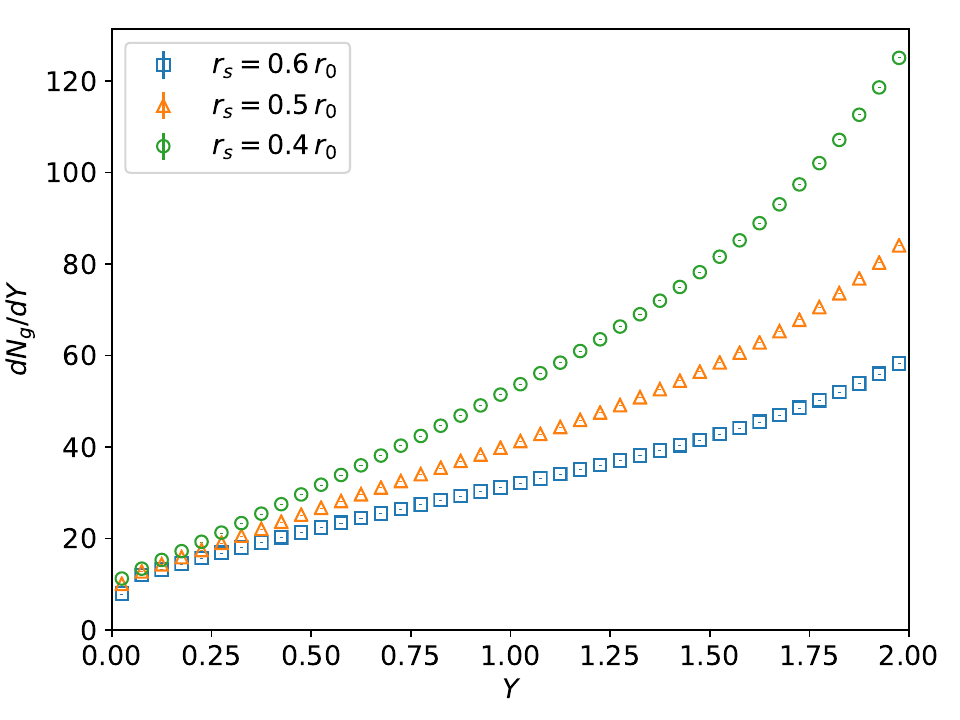}
\includegraphics[width=\linewidth]{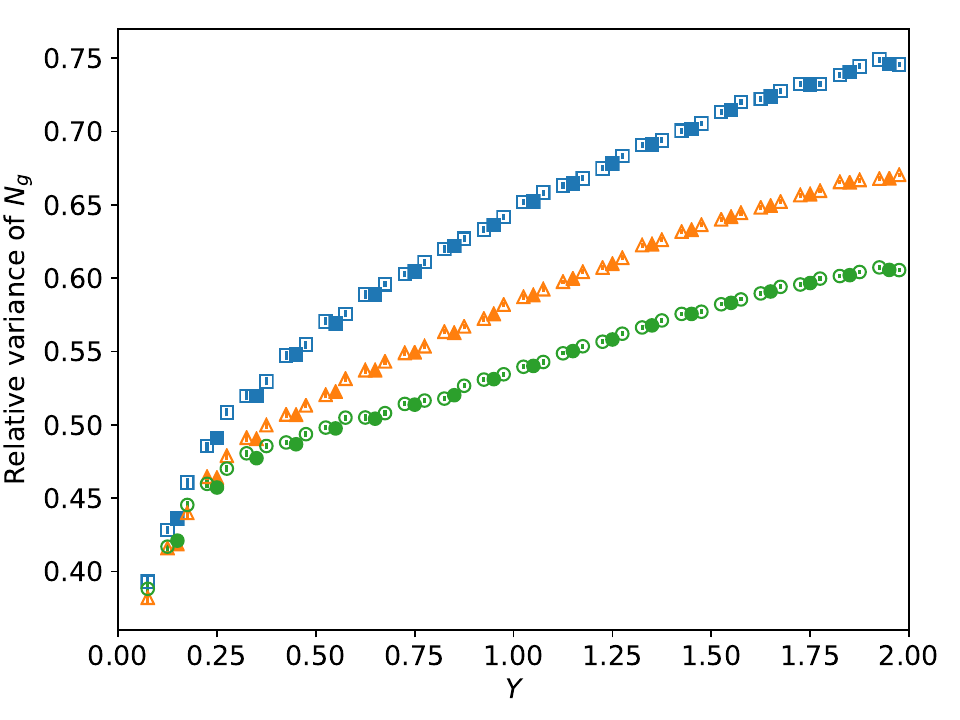}
\end{center}
\caption{(Color online)
  Rapidity dependence of the average multiplicity per unit rapidity (top) and of the relative variance (bottom) in the QCD dipole model, for three different values of the saturation scale $r_s$.
  The relative variance is defined as the variance divided by the square of the mean, and corresponds to the diagonal elements of the relative covariance matrix (\ref{relcovariance}). 
  It has been evaluated with two different bin sizes: 0.05 (empty markers) and 0.1 (filled markers).
  Results are independent of the bin size, as anticipated~\cite{Pruneau:2002yf} when no short-range correlations are present. 
  The rapidity variable $Y$ is defined by Eq.~(\ref{defY}).  
}
\label{fig:dipole}
\end{figure}

\begin{figure}[ht]
\begin{center}
\includegraphics[width=\linewidth]{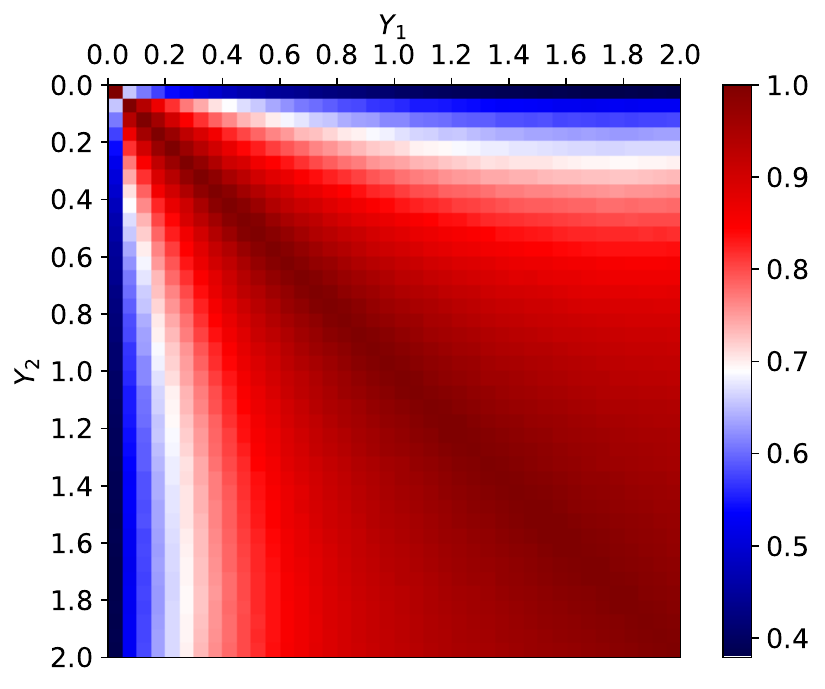}
\end{center}
\caption{(Color online)
  Rapidity correlations in the QCD dipole model. 
  The quantity plotted it the Pearson correlation coefficient (\ref{pearson}) as a function of $Y_1$ and $Y_2$.
  It is plotted in the form of a matrix, so that the vertical axis is descending instead of ascending. 
  The saturation radius is $r_s=0.5\, r_0$.
}
\label{fig:dipolepearson}
\end{figure}

\subsection{Multiplicity fluctuations in the QCD dipole model}
\label{s:dipole}

In order to figure out how the observed phenomena relate with the underlying fundamental QCD dynamics, we now investigate a simple model of gluon production in the scattering of a hadron off a large nucleus, that  can be derived from QCD in an asymptotic limit appropriate to very high energies~\cite{Kovchegov:1998bi,Mueller:2016xti,Liou:2016mfr}.
The hadron in its ground state is modeled as a quark-antiquark color singlet of fixed size $r_0$, referred to as a color dipole.
As for the nucleus, it is solely characterized by a single size $r_s$, called the saturation radius, related to the saturation momentum by $Q_s\sim 1/r_s$. Physically, $r_s$ corresponds to the size of color-singlet domains inside the nucleus in its rest frame. 
The value of $r_s$ is fixed throughout the calculation. 
Therefore, the dipole traverses a slab of nuclear matter of fixed length, similar to a proton-nucleus collisions at $b=0$.

This model is too simple to be quantitatively compared with proton-nucleus data, but we can at least investigate whether the same qualitative trends are present.
We can evaluate the observables studied above, namely: the rapidity dependence of the mean multiplicity and of the variance, and the rapidity correlation. 

Let us briefly describe the model. 
We first need to compute the quantum fluctuations in the wave functions of the colliding objects.  
The density fluctuations in the nucleus are neglected. 
The trick of the dipole model is to start from the rest frame of the nucleus. 
When the process is viewed in this frame, all the fluctuations can be ascribed to the hadron.
The gluons that are emitted in the collision are already present in the wave function of the hadron. 
The ones that eventually go to the final state are liberated thanks to the energy transfer induced by the interaction.

We now explain how the gluon content of the simplified hadron is determined. 
This content depends on the rapidity at which one observes it.
The larger the rapidity, the more gluons one sees.\footnote{We neglect the production of quark-antiquark pairs which becomes negligible as rapidity increases.}
In the framework of the QCD color dipole model \cite{Mueller:1993rr}, the increase of the number of gluons as a function of rapidity is modeled as a Markovian process.
Starting with a quark-antiquark pair at rest, of size vector $\vec r_0$ in the plane transverse to the collision axis, and increasing its rapidity by the infinitesimal quantity $dy$, an additional gluon may be found at the transverse position $\vec r_1$ (up to $d^2 \vec r_1$) with respect to the antiquark with the probability
\begin{equation}
  \frac{\alpha_sN_c}{\pi}dy\frac{r_0^2}{r_1^2(\vec r_0-\vec r_1)^2} \frac{d^2 \vec r_1}{2\pi},
  \label{eq:dipole-splitting}
\end{equation}
where $\alpha_s\ll 1$ is the strong coupling constant, and $N_c$ is the number of colors. 
Note that each gluon is assigned a transverse position through this equation. 
This gluon emission can be viewed as the branching of one dipole into two dipoles of respective size vectors $\vec r_1$ and $\vec r_0-\vec r_1$, the endpoints of which coincide with the antiquark and the gluon, and with the quark and the gluon, respectively.\footnote{The identification of the quark-antiquark-gluon system with a pair of dipoles actually holds true in the large-$N_c$ limit.}
Upon a further increase of the rapidity, each of these two dipoles may split independently into two other dipoles, with the same probability function (\ref{eq:dipole-splitting}), up to the substitution of the size vectors. This further branching corresponds to a second gluon emission, off the quark, the antiquark, or off the first gluon.
Thus the evolution of the gluon content as a function of rapidity is generated by a Markov chain whose kernel is given by Eq.~(\ref{eq:dipole-splitting}).

The expression of this kernel follows from a perturbative QCD calculation (see e.g. \cite{Kovchegov:2012mbw} for a review). 
The fact that hadrons have a finite transverse size is an essential property for our study, but this property is essentially nonperturbative and must be added by hand.
We account for it by supplementing the branching probability with the multiplicative cutoff function 
\begin{equation}
e^{-r_1^2/(2R_\text{IR}^2)}\times e^{-(\vec r_0-\vec r_1)^2/(2R_\text{IR}^2)},
\end{equation}
designed to strongly suppress branchings into dipoles of sizes larger than $R_\text{IR}$~\cite{Liou:2016mfr}, where $R_\text{IR}$ is an infrared cutoff of the order of 1~fm.

The gluonic content is evaluated up to the rapidity of the nucleus, where the scattering occurs.
These gluons are virtual, in the sense that they are quantum fluctuations with a finite lifetime.
In order for a gluon to become real, it needs to be put on-shell through a transfer of energy in the scattering process. 
Inspired by the McLerran-Venugopalan semi-classical approximation~\cite{McLerran:1993ni}, we assume that the dipoles which scatter are those of size $r$ larger than $r_s$.  
The gluons which go to the final state are those which coincide with the endpoints of these dipoles, as well as the gluons which belong to the set of their ancestor dipoles.
The rapidity of a gluon is defined as the rapidity at which is was produced. 

This whole picture can be justified from first principles in the so-called ``double-logarithmic approximation'' if the parameters are strongly ordered according to $r_s\ll r_0\ll R_\text{IR}$. 
In practice, we apply it by choosing parameters relevant for current experiments, where there is no such strong ordering.
The infrared cutoff $R_\text{IR}$ is chosen slightly  larger than the initial dipole size $r_0$, namely $1.5\, r_0$.
The saturation radius of the nucleus $r_s$ should be of the order of $A^{-1/6}r_0$~\cite{McLerran:1993ni}, where $r_0$ is a typical hadronic size and $A=208$ is the atomic number of the nucleus.
We implement three different values: $r_s=0.4\, r_0$, $0.5\, r_0$ and $0.6\, r_0$.

The numerical implementation of the model is described in Ref. \cite{Domine:2018myf}.
The rapidity variable $Y$ in our calculation is related to the physical rapidity by
\begin{equation}
  \label{defY}
Y=\frac{\alpha_sN_c}{\pi}(y-y_h),
\end{equation}
where $y_h$ is the rapidity of the hadron. 
The gluon content is evaluated for $y>y_h$, corresponding to positive $Y$. 
The multiplicative factor in Eq.~(\ref{defY}) is motivated by Eq.~(\ref{eq:dipole-splitting}), which implies that the results are independent of $\alpha_s$ when plotted as a function of $Y$.  
The maximum value of $Y$ in our simulation, corresponding to the rapidity of the nucleus, is  $Y=2$.
This matches the rapidity range of the LHC, which is $\sim 17$, if $\alpha_s\sim 0.12$.
The reason why we pick such a small value for the strong coupling is that the leading-order dipole rapidity evolution, as encoded in the kernel~(\ref{eq:dipole-splitting}), is known to be too fast: setting a value of the coupling smaller than what would be realistic effectively slows down the latter, and makes the gluon number density closer to that expected in the data.

The probability distribution of the number of gluons is shown in Appendix~\ref{s:gluons} for two different rapidity windows.
It is qualitatively similar to the distribution of multiplicity in Angantyr simulations and in experiment.

Figure~\ref{fig:dipole} presents the average gluon density $dN_g/dY$ and its relative variance as a function of the rapidity variable $Y$.
As in Sec.~\ref{s:relativecovariance}, the relative variance is defined as the variance divided by the square of the mean. 
Both the mean multiplicity and the relative variance increase as a function of rapidity, similar to the observation in p+Pb collisions.
The increase of the relative variance as a function of rapidity can be put down to the branching process (\ref{eq:dipole-splitting}).
The number of branchings necessary to produce a gluon typically increases as a function of its rapidity.
Each process is random, and increasing the number of processes also increases the randomness, hence the increase in the relative variance.
This increase is more pronounced for larger values of $r_s$, corresponding to smaller values of the saturation momentum $Q_s$. 

Note that the absolute value of the relative variance in Fig.~\ref{fig:dipole} is much larger than in data and in Angantyr simulations of p+Pb collisions (compare with Eqs.~(\ref{atlascov}) and (\ref{angantyrcov})).
This discrepancy can be ascribed to the fact that a proton does not reduce to a quark-antiquark pair.
Including more components typically reduces the relative variance. 
For a proton consisting of $N$ independent dipoles of identical sizes, for instance, the relative variance would be smaller by a factor $N$. 

Finally, Fig.~\ref{fig:dipolepearson} depicts the rapidity correlation.
We plot the Pearson correlation coefficient, defined as
\begin{equation}
\label{pearson}
r(Y_1,Y_2)\equiv\frac{\sigma(Y_1,Y_2)}{\left(\sigma(Y_1,Y_1)\sigma(Y_2,Y_2)\right)^{1/2}},
\end{equation}
where $\sigma(Y_1,Y_2)$ is the relative covariance matrix.
It gradually deviates from unity as the difference between the two rapidities increases.
This phenomenon is responsible for the rapidity decorrelation, which has been much studied first in the context of anisotropic flow~\cite{CMS:2015xmx} and more recently for multiplicity fluctuations~\cite{Rohrmoser:2019xis,Jia:2020tvb}. 
In the dipole model, the correlation remains very strong even at large relative rapidities.

\subsection{Multiplicity fluctuations in the fluctuating string model}
\label{s:string}

For the sake of comparison, we briefly discuss a different model, the fluctuating-string model, which has been successful in describing data on longitudinal correlations~\cite{Bozek:2015bna,Broniowski:2015oif,Rohrmoser:2019xis} but, unlike the color dipole model, fails to predict the increase of the relative variance with rapidity, at least in its simplest version.  

In string models, hadrons are produced by the fragmentation of strings. 
In the simplest version of the fluctuating-string model, each string produces a uniform rapidity density, over some interval in rapidity. 
But the end points of the interval are allowed to fluctuate event by event, which gives rise to multiplicity fluctuations. 
In the simplest version of the model, only one end of the string fluctuates (Fig.~1 of Ref.~\cite{Bozek:2015bna}), that on the proton side, and the location of the end point is distributed uniformly in rapidity. 
The mean multiplicity and the covariance matrix can be calculated analytically (Appendix B1 of Ref.~\cite{Broniowski:2015oif}). 
We denote by $y_p$ the rapidity of the proton and by $y_{Pb}$ that of the nucleus. 
For a single string, the multiplicity density increases linearly with rapidity~\cite{Bialas:2004su,Bzdak:2012tp} for $y_p<y<y_{Pb}$: 
\begin{equation}
\label{multdipole}
\frac{dN}{dy}\propto y-y_p,
\end{equation} 
and the relative covariance matrix is given by 
\begin{equation}
\label{covdipole}
\sigma(y_1,y_2)=\frac{y_{Pb}-y_2}{y_2-y_p},
\end{equation} 
where the rapidities have been ordered according to $y_1\le y_2$.
The relative variance corresponds to the limiting case $y_1=y_2=y$: 
\begin{equation}
\label{vardipole}
\sigma(y,y)=\frac{y_{Pb}-y}{y-y_p}.
\end{equation}  
It {\it decreases\/} as a function of $y$. 
This example illustrates that the increase seen in ATLAS data, which is reproduced both by Angantyr and by the color dipole model, is not a trivial phenomenon. 
For $N$ independent strings, the relative variance is smaller by a factor $N$. 
We have not investigated whether more sophisticated versions of the fluctuating string model, including double-end fluctuations and fluctuations in the string tension~\cite{Bialas:1999zg,Broniowski:2015oif}, may fix this wrong rapidity dependence. 
 
We also derive for completeness the Pearson correlation coefficient describing the rapidity decorrelation.
Inserting Eq.~(\ref{covdipole}) into (\ref{pearson}), one obtains:
\begin{equation}
  \label{pearsonstring}
r(y_1,y_2)=\sqrt{\frac{(y_1-y_p)(y_{Pb}-y_2)}{(y_2-y_p)(y_{Pb}-y_1)}}, 
\end{equation}
where we again order rapidities according to $y_1\le y_2$.
The Pearson correlation coefficient is typically smaller than in the color dipole model (Fig.~\ref{fig:dipolepearson}), corresponding to a stronger rapidity decorrelation.

\subsection{Discussion of related analyses}
\label{s:discussion}

We finally discuss how our findings relate to previous analyses:
The analysis of long-range multiplicity correlations carried out by ATLAS in Ref.~\cite{ATLAS:2016rbh}, and the centrality dependence of the rapidity spectrum $dN/dy$ in p+Pb collisions~\cite{ALICE:2014xsp,ATLAS:2015hkr}. 

The ATLAS collaboration has measured the relative covariance matrix of the multiplicity distribution $\sigma(\eta_1,\eta_2)$, defined as in Eq.~(\ref{relcovariance}), in narrow bins of $\eta$, in the range $-2.5<\eta<2.5$.
The solid symbols (labeled $|\eta_-|<0.1$) in the middle panel of Fig.11 of Ref.~\cite{ATLAS:2016rbh} represent the variation of $1+\sigma(\eta,\eta)$ as a function of $\eta_+=2\eta$ in p+Pb collisions at $5.02$~TeV.
As $\eta$ increases, the relative variance $\sigma(\eta,\eta)$ first decreases, down to a minimum value compatible with $0$ at $\eta=0$, then increases up to a maximum value of $\sim 0.009$ at $\eta=2.5$. 
At first sight, these results seem incompatible with our claim that the relative variance increases monotonically as a function of $\eta$. 
In addition, the order of magnitude of $\sigma(\eta,\eta)$ is smaller by an order of magnitude than our values in Eq.~(\ref{atlascov}).

The only major difference lies in the definition of the sample of events over which the variance is evaluated. 
In this paper, the event sample consists of all collisions at $b=0$. 
In Ref.~\cite{ATLAS:2016rbh}, it consists of events with a fixed total multiplicity (more precisely, events where the multiplicity in $|\eta|<2.5$ lies within a narrow bin). 
If one fixes the integral of $dN/dy$, the only remaining source of multiplicity fluctuations is a fluctuation in the shape of $dN/dy$~\cite{Lappi:2009vb}.
The largest source of such fluctuation is the forward-backward asymmetry\cite{Bialas:2011bz}, which naturally generates a variance proportional to $|\eta|^2$ (referred to as ``butterfly'' fluctuations in Ref.~\cite{Bzdak:2012tp}). 

We finally discuss the centrality dependence of $dN/dy$~\cite{ALICE:2014xsp,ATLAS:2015hkr}. 
One observes experimentally that as the collision becomes more central, the asymmetry between the p-going side and the Pb-going side becomes more and more pronounced, down to small centrality percentiles~\cite{ATLAS:2015hkr}. 
We argue that this is a natural consequence of the fact that the relative variance increases towards the Pb-going side, as shown in Sec.~\ref{s:relativecovariance}, and that the multiplicity is strongly correlated across the rapidity range, as exemplified by the large value of $\sigma_{12}$.
In experiment, centrality is defined according to multiplicity in some rapidity window. 
More central is defined as higher multiplicity.
With this definition, the variation of impact parameter is essentially irrelevant for the 10\% most central collisions, and one may consider for simplicity that they are all at $b=0$. 

In the limit where the rapidity correlation is maximal, corresponding to a Pearson correlation coefficient (\ref{pearson}) equal to unity, the deviation of $dN/dy$ from its mean value is proportional to the square root of the variance for all $y$:
\begin{equation}
  \label{fluctuatingdndy}
  \frac{dN}{dy}
  =\left\langle\frac{dN}{dy}\right\rangle\left(1+X\sqrt{\sigma(y,y)}\right),
\end{equation}
where $X$ is a random number with zero mean and unit variance which characterizes the amplitude of the multiplicity fluctuation. 
More central events correspond to events with higher $X$.
Since both $\langle dN/dy\rangle$ and  $\sigma(y,y)$ increase with $y$, the asymmetry between the p-going side $y<0$ and the Pb-going side $y>0$ becomes more and more pronounced as $X$ increases.
This qualitatively explains the trend observed experimentally, that the asymmetry increases down to arbitrarily small centrality percentiles.
We intend to carry out a more quantitative study in a forthcoming publication.

\section{Conclusion}

We have introduced a method to reconstruct the averages, standard deviations and covariances of multiplicities in p+Pb collisions at $b=0$, using as input data for minimum-bias collisions.
This method is particularly robust in the sense that it does not rely on any specific model of the collision. 
We manage to reconstruct average values with $\sim 3\%$ accuracy, and standard deviations with $\sim 10\%$ accuracy.
The method has been applied to ATLAS data at $5.02$~TeV. 

Isolating $b=0$ collisions is a crucial improvement for the study of fluctuations.
In minimum-bias data, the dominant source of fluctuations is the decrease of the length of nuclear matter traversed by the proton as $b$ increases.
By selecting $b=0$ collisions, we eliminate this trivial source of fluctuations, and we isolate quantum fluctuations. 
Quantum fluctuations are present in both proton and nuclear wave functions, and also in the collision process. 
These various sources cannot be separated, but it seems natural that the fluctuations are larger in the smaller system, namely, the proton. 
This implies that proton-nucleus collisions have the ability to constrain the description of the proton substructure. 
We have illustrated this by comparing ATLAS data with calculations with the Angantyr model, which turns out to overestimate multiplicity fluctuations. 
We have also shown that the relative multiplicity  fluctuation increases as a function of rapidity. 
This is a non-trivial effect: The fluctuating string model, in its simplest version, predicts a decrease instead. 
We have argued on the basis of the QCD dipole model that this increase, which originates from the soft gluon cascade, is sensitive to the saturation scale.

\section*{Acknowledgements}
JYO thanks Kristjan Gulbrandsen for explaining how to build the correlated gamma fluctuation kernel, which was the starting point of this paper. 
He also thanks Jean-Paul Blaizot, Fran\c cois Gelis, Matt Luzum, Sasha Milov, and Al Mueller for discussions. 
We thank Jiangyong Jia for sending us ATLAS data, and Gr\'egory Soyez for help with \textsc{Pythia}~8. 
This work is supported in part by the GLUODYNAMICS project funded by the ``P2IO LabEx (ANR-10-LABX-0038)'' in the framework ``Investissements d’Avenir'' (ANR-11-IDEX-0003-01) managed by the Agence Nationale de la Recherche (ANR), France. 
PC gratefully acknowledges funding from the Knut and Alice Wallenberg Foundation (the CLASH project).

\appendix
\section{Mapping Gaussian to gamma}
\label{s:mapping}

The one-dimensional Gaussian distribution is
\begin{equation}
\label{gaussian}
P_G(N')=\frac{1}{\sigma\sqrt{2\pi}} \exp\left(-\frac{(N'-\bar N')^2}{2\sigma^2}\right). 
\end{equation}
Its cumulative distribution is 
\begin{eqnarray}
\label{gaussiancumul}
F_G(N')&\equiv &\int_{-\infty}^{N'}P_G(x)dx\cr
&=&\frac{1}{2}\left[1+{\rm erf}\left(\frac{N'-\bar N'}{\sigma\sqrt{2}}\right)\right].
\end{eqnarray}
The gamma distribution is 
\begin{equation}
\label{gamma}
P_\gamma(N)=\frac{1}{\Gamma(k)\theta^k}N^{k-1} e^{-N/\theta},
\end{equation}
with $N>0$. 
Its mean and standard deviation are
\begin{eqnarray}
  \label{meanvariance}
\bar  N &=&\theta k\cr
\sigma &=&\theta\sqrt{k}
\end{eqnarray}
These two equations define, for each gamma distribution, the Gaussian distribution (\ref{gaussian}) with the same mean and variance. 
The cumulative distribution is:
\begin{eqnarray}
\label{gammacumul}
F_\gamma(N)&\equiv &\int_0^{N}P_\gamma(x)dx\cr
&=&\frac{1}{\Gamma(k)}\gamma\left(k,\frac{N}{\theta}\right),
\end{eqnarray}
where $\gamma(s,x)$ denotes the lower incomplete gamma function. 
Eq.~(\ref{mapping}) can be rewritten as:
\begin{equation}
F_\gamma(N)=F_G(N'). 
\end{equation}
The mapping between $N$ and $N'$ is obtained by solving numerically this equation, with the explicit expressions (\ref{gaussiancumul}) and (\ref{gammacumul}). 
It is the bottleneck of our reconstruction procedure, which is much slower than that of Ref.~\cite{Yousefnia:2021cup} for this reason. 

\begin{figure}[ht]
\begin{center}
\includegraphics[width=\linewidth]{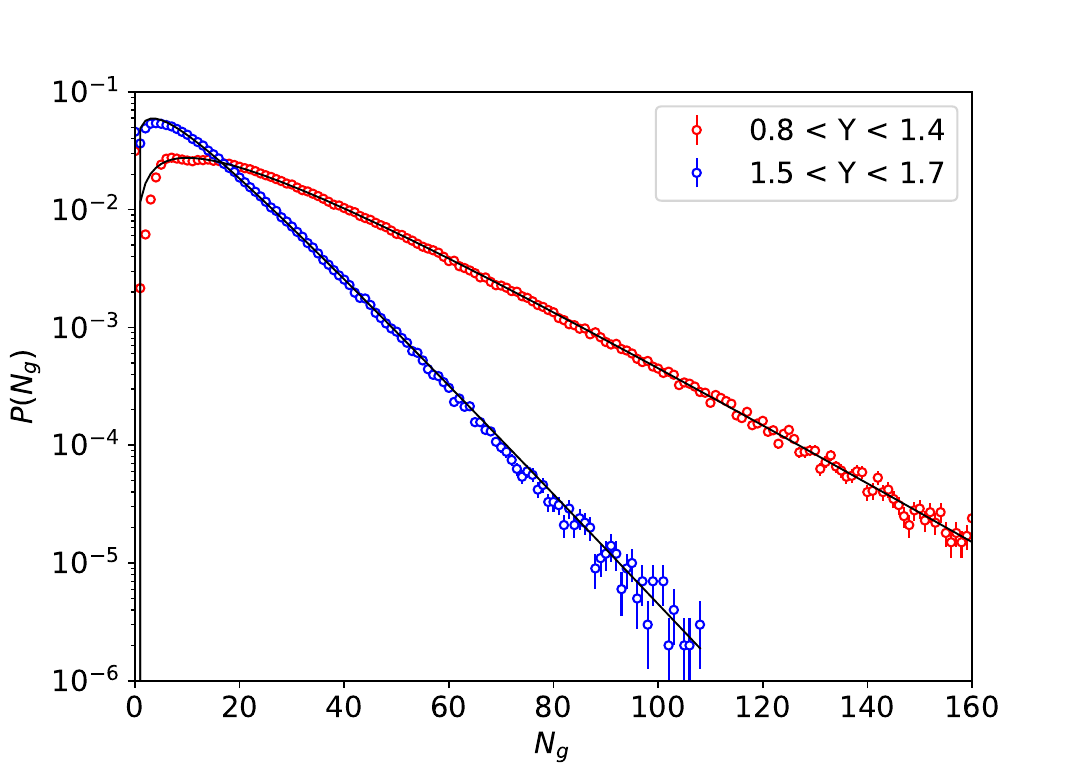}
\end{center}
\caption{(Color online)
  Distribution of the number of gluons in two different rapidity intervals.
  Symbols: Monte Carlo simulation with the QCD dipole model.  
  Lines: gamma distributions with the same mean and variance as the simulation. 
  }
\label{fig:gluon1d}
\end{figure}

\section{$90\%$ confidence ellipse} 
\label{s:confidence}

Here we recall the definition of the $90\%$ confidence ellipse for a generic two-dimensional Gaussian distribution. 
We start with a symmetric two-dimensional Gaussian distribution centered at the origin, with unit width, for two variables $t_1$ and $t_2$: 
\begin{equation}
p(t_1,t_2)= \frac{1}{2\pi} \exp\left(-\frac{t_1^2+t_2^2}{2}\right)
  =\frac{1}{2\pi}\exp\left(-\frac{r^2}{2}\right),
\end{equation}
where $r^2=t_1^2+t_2^2$. 
The probability $f$ that $r<r_0$ is obtained by integrating over the disk of radius $r_0$:
\begin{equation}
  \label{fconfidence}
  f=\int_{r<r_0} p(t_1,t_2)dt_1dt_2
  =1-\exp\left(-\frac{r_0^2}{2}\right).
 \end{equation}
The 90\% confidence circle is the circle which contains 90\% of the probability, that is, the circle for which $f=0.9$.
Its radius $r_0$ is obtained by inverting Eq.~(\ref{fconfidence}):
\begin{equation}
\label{valuer0}
  r_0=\sqrt{-2\ln(1-f)}.
\end{equation}

This can be generalized to an arbitrary Gaussian distribution by carrying out the following change of variables:
\begin{equation}
t_1^2+t_2^2\rightarrow \sum_{j,k}(x_j-\bar x_j) \Sigma^{-1}_{jk} (x_k-\bar x_k),
\end{equation}
where the new variables are $(x_1,x_2)$.
The above equation can be rewritten in matrix form as
\begin{equation}
  ^tTT=(^tX-{^{t}{\bar X}})\Sigma^{-1}(X-\bar X)
\end{equation}
where $T$ and $X$ are column vectors, and $^tT$ and $^tX$ their transpose (line vectors). 
We then solve this equation to express $X$ as a function of $T$: 
\begin{equation}
\label{variablechange}
X=\bar X+\Sigma^{1/2}T, 
\end{equation}
where $\Sigma^{1/2}$ is the  square root of the positive semi-definite matrix $\Sigma$ whose eigenvalues are all positive:
\begin{equation}
  \Sigma^{1/2}=\frac{1}{\sqrt{{\rm Tr}\Sigma+2\sqrt{|\Sigma|}}}
  \left(\Sigma+I\sqrt{|\Sigma|}\right),
\end{equation}
where ${\rm Tr}\Sigma$ denotes the trace, $|\Sigma|$ the determinant, and 
$I$ the $2\times 2$ identity matrix.
Through Eq.~(\ref{variablechange}), the confidence circle of radius $r_0$ is mapped into an ellipse centered at $\bar X$. 
This is the ellipse depicted in Fig.1 of Ref.~\cite{Yousefnia:2021cup} (with $f=0.99$). 
In the case of the correlated gamma distribution, one must at the end carry the transformation from the Gaussian variables to the gamma variables, as explained in Sec.~\ref{s:mapping}.
Due to the nonlinearity of this transformation, the $90\%$ confidence curve depicted in Fig.~\ref{fig:b=0pPb} (see also Figs.~\ref{fig:MBpPb} and \ref{fig:ATLASpPb}) is no longer an ellipse, but it still contains 90\% of the probability by construction. 

\begin{figure*}[ht]
\begin{center}
\includegraphics[width=.49\linewidth]{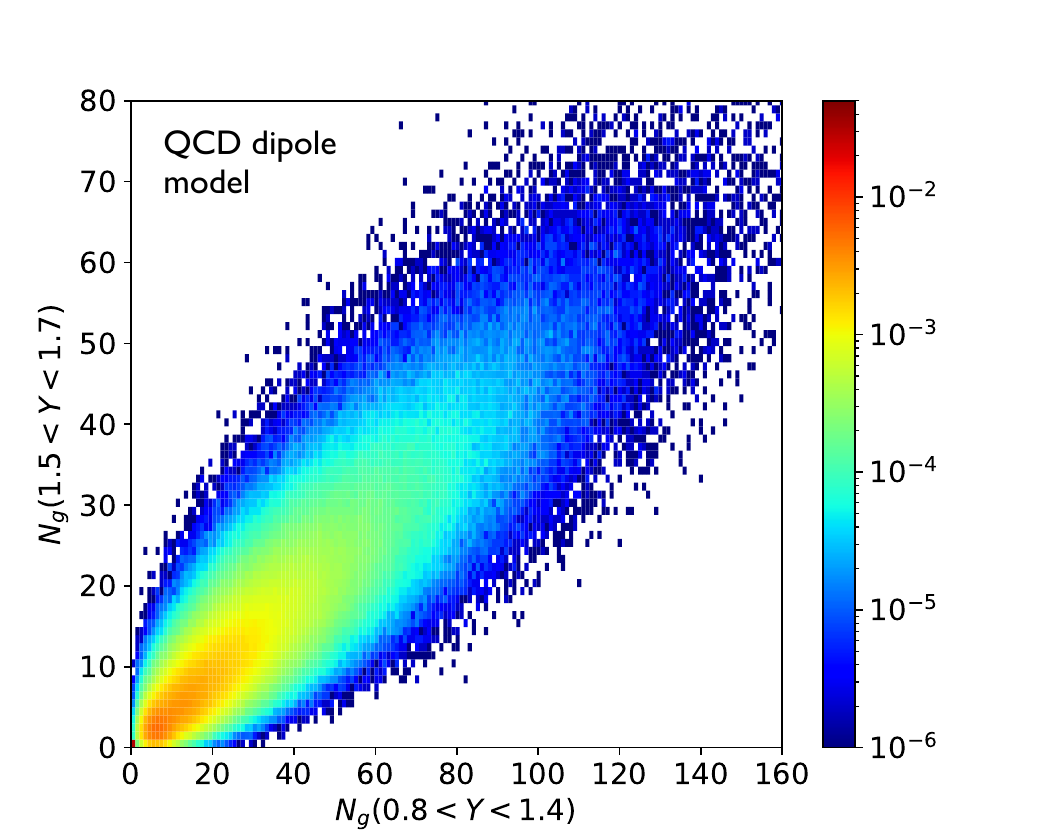}
\includegraphics[width=.49\linewidth]{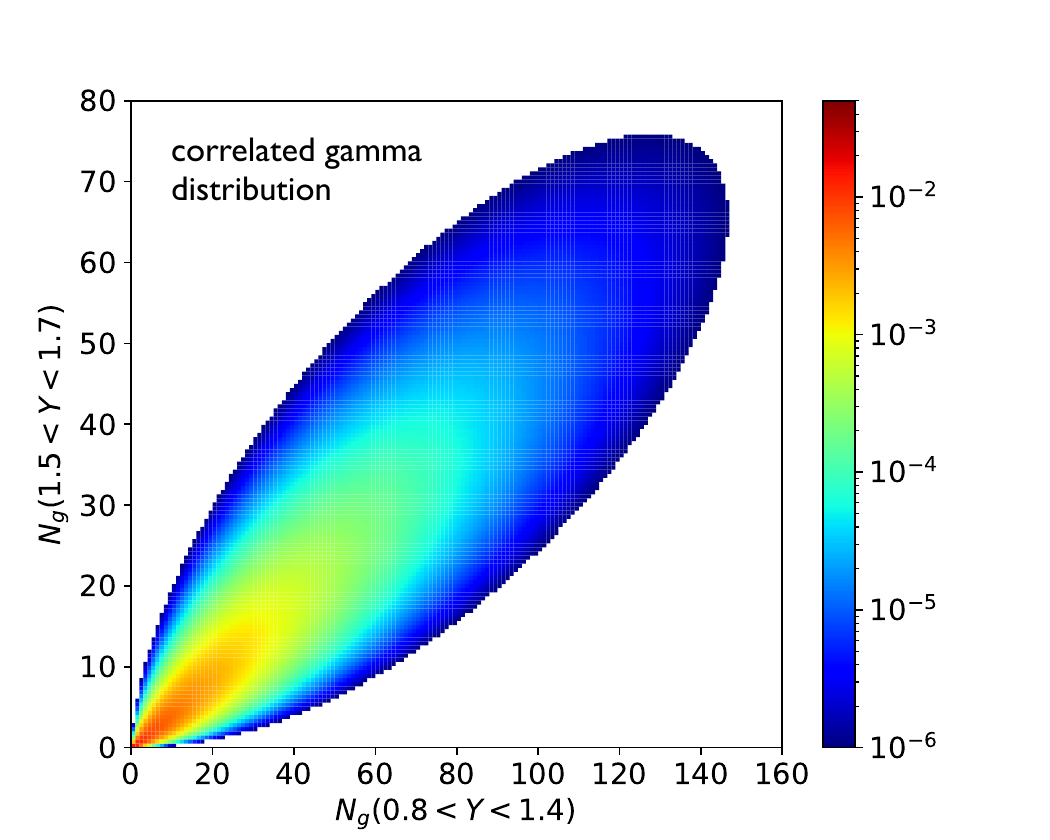}
\end{center}
\caption{(Color online)
  Left: joint distribution of the number of gluons in the two rapidity intervals.
Right: correlated gamma distribution with the same means, variances and covariance. }
\label{fig:gluon2d}
\end{figure*}

\section{Gluon number fluctuations in the QCD dipole model}
\label{s:gluons}

We present numerical results on the distribution of gluon numbers in the Monte Carlo implementation of the QCD dipole model presented in Sec.~\ref{s:dipole}, in order to illustrate their similarity with distributions of multiplicity and transverse energy shown in the main body of the paper. 
We generate $10^6$ events with the saturation radius $r_s$ set to $0.5\, r_0$.
For each event, we compute the gluon number in two different rapidity intervals, whose disposition and size are similar to the pseudorapidity intervals in which ATLAS measures the charged multiplicity $N_{ch}$ and the forward transverse energy $E_T^{\rm Pb}$.
The distributions of these two numbers are displayed in Fig.~\ref{fig:gluon1d}. 
The number of gluons is smaller in the interval at the larger rapidity, despite the fact that $dN/dY$ increases with $Y$ (Fig.~\ref{fig:dipole}), because the interval is narrower. 
The distributions are very close to gamma distributions, shown as lines in Fig.~\ref{fig:gluon1d}. 
The rms errors, defined by Eq.~(\ref{rmserror}) are $16.0\%$ for the rapidity interval $0.8<Y<1.4$, and  $7.5\%$ for the interval $1.5<Y<1.7$. 

The joint distribution is displayed in the left panel of Fig.~\ref{fig:gluon2d}. 
The right panel displays the correlated gamma distribution with the same mean and covariance matrix. 
Both distributions look very similar by eye. 
The rms error as defined from the Kullback-Leibler divergence is in fact rather large, $43.5\%$. 
But it is largely dominated by the first few bins, where the fit overestimates the simulation by a large factor.
If we had excluded the first few bins, as in fits to $N_{ch}$ and $E_T^{\rm Pb}$ distributions shown in this paper, the rms error would be much smaller.

\end{document}